\documentclass[aps, a4paper,superscriptaddress, nofootinbib, showpacs, twocolumn, 10pt]{revtex4}
\usepackage{amsmath,amssymb, bm, natbib}
\usepackage{dsfont}
\usepackage{epsfig}
\usepackage{graphicx}
\usepackage{slashed}
\usepackage{color}
\newcommand{\be}{\begin{equation}}
\newcommand{\ee}{\end{equation}}
\newcommand{\wt}{\widetilde}
\newcommand{\beq}{\begin{equation}}
\newcommand{\eeq}{\end{equation}}
\newcommand{\bea}{\begin{eqnarray}}
\newcommand{\eea}{\end{eqnarray}}
\newcommand{\ba}{\begin{align}}
\newcommand{\ea}{\end{align}}
\newcommand{\bfig}{\begin{figure}}
\newcommand{\efig}{\end{figure}}

\newcommand{\as}{\alpha_s}
\newcommand{\wh}{\widehat}
\newcommand{\nn}{\nonumber}

\newcommand{\eqn}[1]{(\ref{#1})}

\begin{document}
~\vspace{1cm}
\title{Perturbative expansion of the QCD Adler function improved by renormalization-group summation and analytic continuation in the Borel plane}
\author{Gauhar Abbas}
\affiliation{Institute of Mathematical Sciences, C.I.T.Campus,
Taramani, Chennai 600 113, India}
\author{B. Ananthanarayan}
\affiliation{Centre for High Energy Physics,
Indian Institute of Science, Bangalore 560 012, India}
\author{Irinel Caprini}
\affiliation{Horia Hulubei National Institute for Physics and Nuclear Engineering,
P.O.B. MG-6, 077125 Magurele, Romania}
\author{Jan Fischer}
\affiliation{Institute of Physics, Academy of Sciences of the Czech Republic, 
CZ-182 21  Prague 8, Czech Republic}
 
\begin{abstract}  
We examine the large-order behaviour of a recently proposed renormalization-group-improved 
expansion of the Adler function in perturbative QCD,
which sums in an analytically closed form the leading logarithms
accessible from renormalization-group invariance. The expansion is first written as an
effective series in
powers of the one-loop coupling, and its leading singularities in the
Borel plane are shown to be identical to those of the standard
``contour-improved" expansion. Applying the technique of conformal
mappings for the analytic continuation in the Borel plane, we define a
class of improved expansions, which implement both the renormalization-group invariance and
the knowledge about the large-order behaviour of the series. Detailed
numerical studies of specific models for the Adler function indicate that
the new expansions have remarkable convergence properties up to high
orders. Using
these expansions for the determination of the strong coupling from the the hadronic width of
the $\tau$ lepton we obtain, with a
conservative estimate of the uncertainty due to the nonperturbative corrections, 
$\alpha_s(M_\tau^2)=  0.3189^{+ 0.0173}_{-0.0151}$,  which translates to $
\alpha_s(M_Z^2)= 0.1184^{+0.0021}_{-0.0018}$.

\end{abstract}
\pacs{12.38.Cy, 13.35.Dx, 11.10.Hi}
\maketitle

\section{Introduction} 

The determination of the QCD coupling constant to  increasing precision
is one of the most important goals of the Standard Model of
particle physics (for a review see \cite{RPP}).  The non-strange hadronic decays of the $\tau$ lepton are an important source of information on this quantity and have been
exploited now for a couple of decades.  The recent calculation of the Adler function to four loops in massless QCD \cite{BCK08} renewed the interest in the extraction of the 
strong coupling $\alpha_s$ at the scale of the $\tau$ mass from the treatment of these processes 
\cite{Davier2008,MaYa,BeJa,CaFi2009,Pich_Manchester,CaFi_Manchester,Pich_Muenchen,Beneke_Muenchen,CaFi2011,DV,Abbas:2012,Boito_update}. In this context, several modifications of the 
perturbative expansion of the relevant observables have been proposed.   The main ambiguities affecting perturbation theory
are related to the implementation of renormalization-group invariance and to the large-order behaviour of the series.
The differences between the specific ways of accounting for these properties are the main source of theoretical error on the extraction of  $\alpha_s(M_\tau^2)$. 

In a recent work, Ref. \cite{Abbas:2012}, we applied to the analysis of the hadronic $\tau$ decay width a method of improving the  perturbative expansions in  QCD by summing the leading 
logarithms accessible from renormalization-group invariance, 
proposed in \cite{MaMi, Maxwell} and developed in \cite{Ahmady1,Ahmady2}. The properties of the new expansion, which has been referred to as  ``improved fixed-order perturbation theory",  in the 
complex energy plane were 
investigated and were compared with those of the  ``contour-improved perturbation theory"
(CIPT), and the standard ``fixed-order perturbation theory" (FOPT). The new
expansion has the advantage of being written in an analytically closed form, while CIPT, the alternative approach of implementing renormalization-group (RG) invariance, requires the
numerical solution of  the renormalization-group equation for the strong coupling. 

It is known that the  perturbative expansions of QCD correlators are divergent  in many physically interesting situations, with the coefficients growing 
as $n!$ at large orders $n$ \cite{Muell,Muell1,Broa,Bene}.  Alternatively, the divergent character of the series is inferred from the fact that the expanded correlators, like the 
Adler function, are singular at the origin of the coupling constant plane \cite{tHooft}.
These problems are  present also in perturbative  QED \cite{Dyson}, whose phenomenological success
is explained by the fact that the fine structure constant is numerically very
small.  By contrast,  for a relatively large coupling like $\alpha_s(M_\tau^2)$ in  QCD  the consequences are nontrivial. 

Special  mathematical techniques for divergent series are available, like
Borel summation, which under certain conditions 
recovers the expanded function from its increasing expansion coefficients. 
For QCD the problem was studied for many years, and it is known that the  conditions  of Borel summability are not satisfied  \cite{Muell,tHooft, Beneke} (see also the 
reviews  \cite{Fischer}). In particular, the Borel transform of the Adler function has  singularities in the Borel plane, known as ultraviolet (UV) and infrared (IR) renormalons, the 
latter producing an  ambiguity in the reconstruction of the function. However, if one adopts a certain prescription (e.g., the principal value prescription), it is possible to  exploit 
the available knowledge on the large-order behaviour of the coefficients for defining a new expansion, in which the divergent pattern is considerably tamed.  Such an
approach was proposed in  \cite{CaFi1998,CaFi2000,CaFi2001} and  developed recently
in \cite{CaFi2009,CaFi_Manchester,CaFi2011}, using techniques of series acceleration based on conformal mappings and ``singularity softening" 
(these techniques were also applied by other authors, for
a discussion and references see \cite{CaFi2011}).  We recall that the method of conformal mapping was introduced and applied in particle physics in \cite{CiFi} for extending 
the convergence region  of an expansion beyond the circle of convergence and  for increasing the convergence rate at points lying inside the circle.  As discussed in Ref. \cite{CaFi2011}, 
the method is not applicable to the 
(formal) perturbative series in powers of  $\alpha_s$  because the expanded correlators  are singular 
at the point of expansion, $\alpha_s=0$, but  can be applied  in the Borel  plane. It leads to a modified perturbative expansion in terms of a new 
set of functions, which have the advantage of resembling  the 
expanded correlator (the Adler function), in particular by sharing its 
known singularities  in the coupling and the Borel complex planes.

In Refs. \cite{CaFi2009,CaFi_Manchester,CaFi2011} the method was applied to the two standard versions of perturbation theory, CIPT and FOPT. 
As argued in 
\cite{CaFi2011}, the new expansions are particularly 
suitable in the contour-improved version, 
since they make simultaneously the RG summation and the Borel large-order summation of 
the Adler function. 
Detailed numerical studies \cite{CaFi2009,CaFi2011} established the good 
convergence properties of the latter expansions for several exact models which 
simulate the known properties of the Adler function.  

In this work, we consider the large-order properties of the expansion discussed in \cite{Abbas:2012}, which we shall
henceforth  refer to as ``renormalization-group-summed" (RGS) expansion.
We investigate the properties of this  scheme in the Borel plane and, using the techniques discussed in \cite{CaFi2009,CaFi_Manchester,CaFi2011}, we define a new class of expansions, which simultaneously implement the renormalization-group and the large-order summation by the analytic continuation  in the Borel plane. We shall 
refer to these as ``Borel and renormalization-group-summed" (BRGS) expansions.

The plan of the paper is as follows:
We briefly review in Sec. \ref{sec:Adler} 
the perturbative expansion of 
the Adler function and its connection to the hadronic decay 
width of $\tau$ lepton. 
In Sec. \ref{sec:RGR}
we review the derivation of the RGS expansion following \cite{Abbas:2012}, and show further that it can be expressed as an
expansion in powers of the one-loop solution of the RG equation for the
coupling, which is needed for rendering this expansion suitable for convergence acceleration.   In Sec. \ref{sec:Borel} we discuss the 
properties of the expansion in the Borel plane, 
and show that it has the same dominant singularities as the CI expansion.
 In Sec. \ref{sec:confmap} we define a set of 
new Borel and RG-improved expansions, by using 
the technique of singularity softening and conformal mappings 
of the Borel plane \cite{CaFi2011}. 
In Sec. \ref{sec:tests} we investigate the properties of the new 
expansions in the complex energy plane and illustrate their convergence for the physical observable relevant for the hadronic width of the $\tau$ lepton, using a class of  models for the Adler function considered  in 
\cite{BeJa, CaFi2009, CaFi2011, DeMa, BeBoJa}. In Sec. \ref{sec:alphas} we report
a new determination of 
$ \alpha_s(M_\tau^2)$ based on the new BRGS expansions and
in Sec. \ref{sec:conc} we summarize our results and  conclusions.

\section{Adler function and the hadronic  $\tau$-decay width}\label{sec:Adler}

The Adler function plays a crucial role in the determination of  $\alpha_s(M_\tau^2)$ 
from hadronic $\tau$ decays.  The method is discussed in the seminal paper \cite{BrNaPi} 
and is reviewed in several recent articles 
\cite{Davier2008, BeJa, Pich_Manchester, Pich_Muenchen}. 

The inclusive character of the total $\tau$ hadronic width makes possible  an accurate
calculation of the ratio  
\beq\label{Rtau}
R_\tau \,\equiv\,\frac{\Gamma[\tau^- \to \nu_\tau {\rm hadrons} \,  ]}{
\Gamma[\tau^- \to \nu_\tau e^- \bar \nu_e ]}.
\eeq
Of interest is the Cabbibo allowed component
$R_{\tau,V/A}$ proceeding either through a 
vector or an axial vector current, which can be expressed theoretically in the 
form:
\begin{equation}
\label{RtauVA}
R_{\tau,V/A} \,=\, \frac{N_c}{2}\,S_{\rm EW}\,|V_{ud}|^2\,\biggl[\,
1 + \delta^{(0)} + \delta_{\rm EW}' + \sum\limits_{D\geq 2} 
\delta_{ud}^{(D)} \,\biggr] \,,
\end{equation}
where $N_c =3$ is the number of quark colours,
$S_{\rm EW}$  and
$\delta_{\rm EW}'$  are electroweak corrections,
$\delta^{(0)}$ is the dominant perturbative QCD correction, and  $\delta_{ud}^{(D)}$ denote quark mass
and higher $D$-dimensional operator corrections (condensate contributions) arising in
the operator product expansion (OPE). The decay width is suitable for the precise extraction of the strong coupling, since the (less-known) higher terms in the OPE bring a very small 
contribution to (\ref{RtauVA}). Therefore, a fairly accurate phenomenological determination of the QCD perturbative part $\delta^{(0)}$ is 
possible \cite{Davier2008, BeJa, Pich_Muenchen, Beneke_Muenchen}.

 The theoretical calculation of $\delta^{(0)}$ is based on unitarity, which implies that the inclusive hadronic decay rate can be written as a weighted integral along the timelike axis of 
the spectral function of a polarization function. As shown in \cite{BrNaPi}, the analytic properties of the polarization function and the Cauchy theorem allow one to write  equivalently 
this quantity  as an integral along a contour in the complex
$s$-plane (chosen for convenience to be the circle $|s|=M_\tau^2$).  After an integration by parts,  the quantity of interest $\delta^{(0)}$ is expressed as the  contour integral: 
\begin{equation}\label{del0}
\delta^{(0)}=\frac{1}{2 \pi i}\!\! \oint\limits_{|s|=M_\tau^2}\!\! \frac{d s}{s}
\left(1- \frac{s}{M_\tau^2}\right)^3 \left(1+\frac{s}{M_\tau^2}\right) \widehat{D}_{\rm pert}(s),
\end{equation}
 where the reduced Adler function
 $\widehat{D}(s)\equiv D^{(1+0)}(s)-1$ is obtained
 by subtracting the dominant term  from the logarithmic derivative of the polarization function, $D^{(1+0)}(s)\equiv - s\, {\rm d}\Pi^{(1+0)}(s)/{\rm d}s$,  where the superscript denotes the 
spin \cite{BrNaPi}. The perturbative expansion of $\widehat{D}(s)$ reads \cite{BeJa}
\begin{equation}
\label{Ds}
 \widehat{D}_{\rm pert}(s) = \sum\limits_{n=1}^\infty \,(a_s(\mu^2))^n\,
\sum\limits_{k=1}^{n} k\, c_{n,k}\, (\ln (-s/\mu^2))^{k-1} \,,
\end{equation}
where $a_s(\mu^2)\equiv \as(\mu^2)/\pi$ is the strong coupling  at the renormalization scale  $\mu$. The leading coefficients $c_{n,1}$ are obtained from Feynman diagrams, the known 
coefficients $c_{n,1}$ calculated to four loops in the $\overline{{\rm MS}}$-renormalization scheme being (see \cite{BCK08} and references therein):
\beq\label{cn1}
c_{1,1}=1,\, c_{2,1}=1.640,\, c_{3,1}=6.371,\, c_{4,1}=49.076.
\eeq
Several estimates for the next coefficient $c_{5,1}$ were made recently \cite{Davier2008, BeJa, Beneke_Muenchen, Pich_Muenchen}.

The remaining coefficients $c_{n,k}$ for $k>1$  are determined from renormalization-group invariance:  the function   $\widehat{D}_{\rm pert}$, calculated in a fixed renormalization 
scheme,  is scale independent and therefore satisfies  the equation 
\begin{equation}\label{eq:rgi}
\mu^2 \frac{\mathrm{d}}{\mathrm{d}\mu^2} \left[  \widehat{D}_{\rm pert}(s) \right] =
0,
\end{equation}
which can be written equivalently as
\begin{equation}\label{pde2}
 \beta(a_s)\frac{\partial \widehat{D}_{\rm pert}}{\partial a_s}-\frac{\partial \widehat{D}_{\rm pert}}{\partial \ln (-s/\mu^2) }  = 0,
\end{equation}
where
\begin{equation}\label{beta}
\beta(a_s) \equiv \mu^2 \frac{\mathrm{d}a_s(\mu^2)}{\mathrm{d}\mu^2} = -(a_s(\mu^2))^2 \sum_{k=0}^\infty \beta_k (a_s(\mu^2)) ^k
\end{equation}
is the  $\beta$ function governing the scale dependence of the coupling.  From (\ref{pde2}) one can express  $c_{n,k}$ for $k>1$ in terms of  $c_{n,1}$ and the coefficients $\beta_j$ of 
the perturbation expansion (\ref{beta}). 
 The known $\beta_j$  coefficients, calculated to four loops in  $\overline{{\rm MS}}$  scheme, are (see \cite{LaRi,Czakon} for the calculation of $\beta_3$ and earlier references):
\beq\label{betaj}
 \beta_0=9/4,\,\, \beta_1=4,\,\, \beta_2= 10.0599,\,\,\beta_3=47.228.
\eeq

In the  ``fixed-order perturbation theory"  calculation of  $\delta^{(0)}$, the choice  $\mu^2=M_\tau^2$ is adopted, when the expansion (\ref{Ds}) reads:
\bea\label{DsFO}
 \widehat{D}_{\rm FOPT}(s)& =&\sum\limits_{n=1}^\infty (a_s(M_\tau^2))^n \, [c_{n,1} \\
 &+& \sum\limits_{k=2}^{n} k\, c_{n,k}\, (\ln (-s/M_\tau^2))^{k-1}].\nonumber
\eea
As remarked in \cite{dLP1}, due to the large imaginary part of  the logarithm  $\ln (-s/ M_\tau^2)$  along the circle $|s|= M_\tau^2$, the series (\ref{DsFO})  is badly behaved, especially near the timelike axis.
The  CIPT  \cite{Pivovarov:1991rh,dLP2} is  defined by the choice $\mu^2=-s$, when (\ref{Ds}) reduces to 
\beq\label{DsCI}
 \widehat{D}_{\rm CIPT}(s)= \sum\limits_{n=1}^\infty  c_{n,1} (a_s(-s))^n
\,,
\eeq
 where the running coupling  $a_s(-s)$  is determined by solving the renormalization-group equation (\ref{beta}) numerically in an iterative way along the circle, starting with the input value $a_s(M_\tau^2)$ at $s=-M_\tau^2 $.  This expansion avoids the appearance of large logarithms along the circle $|s|= M_\tau^2$. 

The expansions (\ref{DsFO}) and (\ref{DsCI}) coincide formally as long as all the terms in the series are
retained. In fact, the coefficients $c_{n,1}$ are known to increase as $n!$, so the series are  divergent. We shall turn to this property in Sec. \ref{sec:Borel}.  If the series are
truncated at some order $N$, the expansions (\ref{DsFO}) and (\ref{DsCI})  differ by terms of order $a_s^{N+1}$, this being the main theoretical  error in the  the determination of  $\alpha_s(M_\tau^2)$ from the measured $\tau$ hadronic width.

\section{Renormalization-Group Summation}\label{sec:RGR}
As proposed in \cite{Ahmady1, Ahmady2}, the  expansion (\ref{Ds})  can be written in the RGS form
\begin{equation}
\label{dseries}
 \widehat{D}_{\rm RGS}(s) =
\sum_{n=1}^\infty (a_s(\mu^2))^n D_n (y),
\end{equation}
 where the functions $D_n(y)$, depending on a single variable
\be\label{y}
 y \equiv 1+\beta_0 a_s(\mu^2) \ln (-s/\mu^2),
\ee are defined as 
\begin{equation}\label{Dn_def}
D_n (y) \equiv \sum_{k=n}^\infty (k-n+1)c_{k, k-n+1} \left(\frac{y-1}{\beta_0}\right)^{k-n}.
\end{equation}
As seen from the definition, the  function $D_1$ sums  the leading logarithms in the series (\ref{Ds}), the function $D_2$ sums the next-to-leading logarithms, and so on.  The attractive 
feature pointed out in \cite{Ahmady1, Ahmady2}, is that these functions can be obtained in a closed analytical form. The derivation is based on renormalization-group invariance: after 
inserting  the expansion \eqn{dseries} into the condition (\ref{eq:rgi}), a straightforward calculation leads to the following system of differential equations for $D_n(y)$, for  $n\ge 1$:
\beq\label{Dk_de}
\beta_0 \frac{\mathrm{d}D_n(y)}{\mathrm{d}y} +  \sum_{\ell = 0}^{n-1} \beta_\ell \left( (y-1)\frac{\mathrm{d}}{\mathrm{d}y} + n - \ell \right) D_{n - \ell}(y)=0,
\eeq
with the initial conditions $D_n (1) = c_{n, 1}$ which follow from (\ref{Dn_def}). 

 The solution of the system (\ref{Dk_de}) can be found iteratively in an analytical form.
The expressions of $D_{n}(y)$ for $n \leq 5 $, written in terms of the coefficients $c_{k,1}$ with $k\leq n$ and $\beta_j$ with $0\le j\le n-1$, are:
\beq\label{D1}
D_1 (y) = \frac {c_{1, 1}} {y} 
\eeq
\beq\label{D2}
D_2 (u) =   \frac{1}{y^2}\left(c_{2, 1} +   c_{1,1}\, d_{2,1} \right), \quad d_{2,1}=- \frac{\beta_1}{\beta_0} \ln y,
\eeq
\beq\label{D3}
D_3(y) =  \frac{1}{y^3} \left(c_{3, 1} +
c_{2, 1}\,d_{3,2} + c_{1,1}\, d_{3,1} \right),
\eeq
\begin{align}
&d_{3,2}= -\frac{2 \beta_{1}}{ \beta_{0}  }\, \ln y \\
&d_{3,1}= -\frac{\beta_ {1}^2}{\beta_{0}^2}  (1-y+ \ln y-\ln^2 y)+ \frac{\beta_2}{ \beta_{0}}
(1-y),\nonumber
\end{align}

\beq\label{D4}
D_4 (u) = \frac{1}{y^4} (c_{4,1}  + c_{3,1} \, d_{4,3} 
 +  c_{2,1} \, d_{4,2}  + c_{1,1} \, d_{4,1} )
\eeq
\begin{widetext}
\begin{align}
& d_{4,3} = -3 \frac{\beta_1}{\beta_{0}}  \ln y \\
&d_{4,2} = -2 \frac{\beta_{2}}{\beta_{0}}\,  (-1 + y) + \frac{\beta_{1}^2}{ \beta_{0}^2}\, (-2 - 2 \ln y + 3 \ln^2 y +  2 y) \nonumber \\
&d_{4,1} = -\frac{\beta_{1}^3}{2 \beta_{0}^3} \left (-5 \ln^2 y
+ 2 \ln^3 y + 4 {\ln y} (-1 + y) + (-1 + y)^2 \right) 
 -  \frac{\beta_{1} \beta_{2}}{ \beta_{0}^2} \left (3 \ln y + y - 
      2 y \ln y  - 
      y^2 \right) - \frac{\beta_{3}}{2\beta_0} \left (-1 + 
      y^2 \right) \nonumber
\end{align}
\beq\label{D5}
D_5 (y) = \frac{1}{y^5}(c_{5,1} + c_{4,1}\, d_{5,4} +c_{3,1}\, d_{5,3}
 +c_{2,1}\,d_{5,2}+ c_{1,1}\, d_{5,1}),
\eeq
\begin{align}\label{d5j}
d_{5,4} &= - 4\, \frac{\beta_{1}}{\beta_0} \ln y, \\
d_{5,3}&=-3\left [\frac{ \beta_{1}^2}{ \beta_{0}^2}\,\left(-2 \ln^2 y + {\ln y} - y + 
        1 \right)  + \frac{ \beta_{2}}{ \beta_{0}}  (y - 1) \right],
\nonumber \\
d_{5,2} &=- \left[\frac{\beta_1^3}{\beta_{0}^3}\, (4 \ln^3 y -  7 \ln^2 y + 
      6 (y - 1) \ln y + (y - 1)^2) 
+ 2 \frac{\beta_{2} \beta_{1}}{\beta_{0}^2} (-y^2 - 3 y \ln y + y + 
      4 \ln y)  + \frac{\beta_3}{\beta_0} (y^2 - 1)\right],
\nonumber \\ 
d_{5,1} &=\frac{\beta_1^4}{6 \beta_{0}^4} \left (6 \ln^4 y - 
      26 \ln^3 y + 9 (2 y - 1) \ln^2 y + 
      6\left (y^2 - 5 y + 4 \right) \ln y + (y - 1)^2 (2 y + 
         7) \right)
\nonumber \\
&-   \frac{\beta_1^2 \beta_2 }{\beta_0^3}  \left (3 (y - 2)
         \ln^2 y + \left (2 y^2 - 5 y + 
         3 \right) {\ln y} + (y - 1)^2 (y + 
         3) \right) 
\nonumber \\
&+ \frac{\beta_{1} \beta_{3}} {6\beta_{0}^2} \left (4 y^3 - 3 y^2 + 6 \ln y\left (y^2 - 2 \right) - 
      1 \right)+ 
    \frac{\beta_{2}^2}{3\beta_{0}^2}  (y - 1)^2 (y + 
         5) - \frac{\beta_{4}}{3\beta_0} (y^3 -1). \nonumber
\end{align}

\end{widetext}
The expressions of $D_n(y)$ for $6\leq n\leq 10$, which depend also on the coefficients $c_{n,1}$ for $6\leq n\leq 10$ and  $\beta_j$ for $5\le j\le 9$,  are given in a somewhat different 
form\footnote{For simplicity, in \cite{Abbas:2012} we presented the expressions obtained by inserting the known numerical  values of  $\beta_j$ for $j\le 3$ from (\ref{betaj}), and 
setting $\beta_j=0$ for $j\ge 4$.} in the Appendix of \cite{Abbas:2012}. 
 In the numerical applications presented in Sec \ref{sec:tests}, we shall use the expressions of $D_n(y)$ up to $n=18$, which can be  obtained easily by solving the system (\ref{Dk_de}) 
with a MATHEMATICA program.

Using Eqs. (\ref{D1}) - (\ref{d5j}) and the expressions of higher $D_n(y)$ derived analytically, we note that these functions can be written as
\beq\label{Dn}
D_n(y)=\frac{1}{y^n}\,\left[c_{n,1} + \sum_{j=1}^{n-1} c_{j,1} d_{n,j}(y)\right],
\eeq
where the coefficients $d_{n,j}$ are functions of $y$, which depend also on the
coefficients $\beta_j$ of the $\beta$-function. 
One can check that $d_{n,j}(y)$ vanish identically for $y=1$ or if 
$\beta_j=0$ for $j\ge 1$.

By inserting (\ref{Dn}) into (\ref{dseries}), we note that the denominators $y^n$ can be combined in each term with the powers of $a_s(\mu^2)$, so that (\ref{dseries}) can be written as
\beq\label{dseriesy}
\wh D_{\rm RGS}(s) = \sum_{n=1}^\infty (\wt a_s(-s))^n \, \left[c_{n,1}+ \sum_{j=1}^{n-1} c_{j,1} d_{n,j}(y)\right],
\eeq
where
\beq\label{atilde}
\wt a_s(-s)=\frac{a_s(\mu^2)}{1+\beta_0 a_s(\mu^2) \ln(-s/\mu^2)}
\eeq
is  the solution of the RG equation (\ref{beta}) to one loop at the scale $-s$, written in terms of $a_s(\mu^2)\equiv \alpha_s(\mu^2)/\pi$. The  terms $c_{n,1}$  in  the 
series (\ref{dseriesy}) yield the all-order summation of the one-loop coupling (\ref{atilde}), while the remaining sums yield the corrections accounting for the higher order terms in the 
expansion of the $\beta$-function.

\section{Borel transform}\label{sec:Borel}
In this section we discuss the properties of the expansion (\ref{dseriesy}) in the Borel plane.  We start by recalling the standard definition \cite{BeJa}  of the Borel transform $B(u)$ of 
the expansion (\ref{DsCI}):
\beq\label{B}
B(u)= \sum_{n=0}^\infty c_{n+1,1}\, \frac{u^n}{\beta_0^n \,n!}.
\eeq
The original function $\wh D_{\rm CIPT}(s)$  is recovered from $B(u)$ by a Laplace-Borel integral. Actually,  the function $ B(u)$ has singularities on the positive axis of the $u$ plane, 
so the Laplace-Borel integral requires a regularization. Following Refs. \cite{BeJa, CaFi2009, CaFi2011} we shall use the principal value ({\rm PV})
prescription. Note that as argued in \cite{CaNe,CaFi3}, the {\rm PV}  prescription  preserves the reality  of the correlators in the  $s$-plane, and
is therefore more consistent  than other prescriptions with the analyticity properties imposed by causality and unitarity.  Thus, we write
\beq\label{eq:LaplaceCI}
 \wh D_{\rm CIPT}(s)=\frac{1}{\beta_0}\,{\rm PV} \,\int\limits_0^\infty  
\exp{\left(\frac{-u}{\beta_0 a_s(-s)}\right)
} \, B(u)\, {\rm d} u.
\eeq
Similarly, we can define the Borel transform $B_{\rm FO}(u,s)$ of the FOPT expansion (\ref{DsFO}). The structure of the coefficients of this expansion  implies that we can 
write $B_{\rm FO}(u,s)$ as:
\beq\label{BFO}
B_{\rm FO}(u,s)=B(u)+ \sum_{n=0}^\infty  \frac{u^n}{\beta_0^n \,n!} \sum\limits_{k=2}^{n+1} k\, c_{n+1,k}\,\left(\ln\frac{-s}{M_\tau^2}\right)^{k-1}.
\eeq
 The function $\wh D_{\rm FOPT}(s)$ is obtained from its Borel transform by
\beq\label{eq:LaplaceFO}
 \wh D_{\rm FOPT}(s)=\frac{1}{\beta_0}\,{\rm PV} \,\int\limits_0^\infty  
\exp{\left(\frac{-u}{\beta_0 a_s(M_\tau^2)}\right)
} \, B_{\rm FO}(u,s)\, {\rm d} u.
\eeq
We introduce now the Borel transform $B_{\rm RGS}(u, y)$ of the expansion (\ref{dseriesy}), which can be written as
\beq\label{BRG}
B_{\rm RGS}(u, y)=B(u) + \sum_{n=0}^\infty\frac{ u^n} {\beta_0^n \,n!} \sum_{j=1}^{n} c_{j,1} d_{n+1,j}(y).
\eeq
The function $\wh D_{\rm RGS}(s)$ is recovered  by the similar Laplace-Borel integral
\beq\label{eq:LaplaceRGS}
 \wh D_{\rm RGS}(s)=\frac{1}{\beta_0}\,{\rm PV} \,\int\limits_0^\infty  
\exp{\left(\frac{-u}{\beta_0 \tilde a_s(-s)}\right)
} \, B_{\rm RGS}(u, y)\, {\rm d} u,
\eeq
written in terms of the one-loop coupling (\ref{atilde}).

As we already mentioned, 
the function $B(u)$ defined in (\ref{B}) has singularities on the real axis in the $u$-plane, namely along the rays $u\ge 2$ and $u\leq -1$  \cite{Muell,Beneke}. Moreover, the
nature of the dominant singularities can be described exactly: they are branch points, near which $B(u)$ behaves, respectively,  as
\beq\label{eq:gammapowers}
 B(u) \sim (1+u)^{-\gamma_{1}},  \quad  \quad B(u)  \sim (1-u/2)^{-\gamma_{2}}, 
\eeq
where the exponents
$\gamma_1$ and $\gamma_2$,  calculated using renormalization-group
invariance, have known positive values  \cite{Muell,BBK,BeJa}:
\begin{equation}\label{eq:gamma12}
\gamma_1 = 1.21,    \quad\quad   \gamma_2 = 2.58 \,. 
\end{equation}
From (\ref{BFO}) and (\ref{BRG}) it follows  that these singularities are present also in 
the Borel transforms $ B_{\rm FO}(u, s)$ and
$ B_{\rm RGS}(u, y)$. In principle, these functions might have also other singularities, due to the additional infinite sums appearing in (\ref{BFO}) and (\ref{BRG}), respectively.
However, as we shall argue below,  the dominant singularities of these functions, {\em i.e.} the singularities  closest to the origin $u=0$,  are those
at $u=-1$ and $u=2$ contained in $B(u)$.

We first present some evidence which results from the inspection of the next-to-leading terms in the expression  (\ref{BRG}) of $B_{\rm RGS}(u, y)$. Thus, from the expressions   
(\ref{D1})-(\ref{d5j}) we note that
\beq\label{eq:dnn-1}
d_{n,n-1}(y)= -(n-1) \,\frac{\beta_1}{\beta_0}\,\ln y.
\eeq
By inserting this expression in (\ref{BRG}), we obtain by a straightforward calculation the contribution to $B_{\rm RGS}(u, y)$ of the term with $j=n$:
\beq\label{eq:B1nn}
B_{\rm RGS}(u, y)|_{j=n}= - \frac{\beta_1}{\beta_0^2}\, u B(u)\, \ln y.
\eeq
Similarly, we note that
\beq\label{eq:dnn-2}
d_{n,n-2}(y) =-(n-2) \xi (y) 
 + \frac{(n-1)(n-2)}{2}\,\frac{\beta_1^2}{\beta_0^2}\,\ln^2\!y, 
\eeq
where
\beq
\xi(y)= \frac{\beta_1^2}{\beta_0^2}\,\ln y + \frac{\beta_1^2 -\beta_0\beta_2}{\beta_0^2}(1-y).
\eeq
Then we obtain by a straightforward calculation the contribution of the term with $j=n-1$ in (\ref{BRG}) as
\bea\label{eq:B1nn-1}
 B_{\rm RGS}(u, y)|_{j=n-1}&=&-\frac{\xi(y)}{\beta_0^2}   \int_0^u u' B(u') {\rm d}u'\nonumber\\
  &+&\frac{\beta_1^2}{2\beta_0^4}\,\ln^2\! y\, u^2 B(u),
\eea
where the integral  is defined along  a contour from the origin to the point $u$, which does not reach the singularities of $B(u)$.

 The next coefficients in the second term of (\ref{BRG}) exhibit a similar pattern: $d_{n,n-l}$ for $1<l<n-1$ contains  polynomials of the variables $(1-y)$ and $\ln y$, of 
degree $(l-1)$   and $l$, respectively, with coefficients depending on $\beta_j$, $n$ and $l$. 
 For instance, the term proportional to $\ln^l\! y$  in $d_{n,n-l}$ has the expression
 \beq\label{eq:dnn-j}
d_{n,n-l} \sim \frac{(-1)^l} {l!} \prod_{k=1}^l (n-k) \frac{\beta_1^l}{\beta_0^l}\,\ln^l\!y,
\eeq
bringing a contribution  to (\ref{BRG}) of the form
\beq
 B_{\rm RGS}(u, y)|_{j=n+1-l} \sim  \frac{(-1)^l} {l!}\frac{\beta_1^l}{\beta_0^{2l}}\!y\,u^l B(u) \,\ln^l\!y.
\eeq
This reproduces (\ref{eq:B1nn}) and the second term in (\ref{eq:B1nn-1}) for $l=1$ and $l=2$, respectively.

The other terms appearing in $d_{n,n-l}$ may contribute also with integrals  of $B(u)$ multiplied by powers of $u$, as in (\ref{eq:B1nn-1}). 
Thus,  in general, the second term in the expression (\ref{BRG}) of $B_{\rm BRG}(u, y)$ is expected to contain either $B(u)$, multiplied by factors which vanish in the 
limit $y\to 1$, or  integrals of  $B(u)$, in which the singularities have the same positions as in (\ref{eq:gammapowers}), but are  milder. Therefore, the dominant 
singularities of the Borel transform (\ref{BRG}) concide with the dominant singularities of the Borel transform $B(u)$ defined in (\ref{B}).

One may  invoke also the general argument that  Mueller  \cite{Muell} used for concluding that the dominant singularities of the Borel transform $B_{\rm FO}(u,s)$ defined 
in (\ref{BFO}) are the same as those of $B(u)$.  The crucial observation is that the positions of the dominant singularities of the Borel transform  are determined from the 
behaviour of the correlators in the limit of a small coupling.\footnote{The connection between the behaviour of the Borel-summed correlators in the $a_s$-plane and the position 
and  nature of the dominant singularities of the Borel transform  in the $u$ plane is given 
explicitly in the case of a finite number of
renormalons in \cite{CaFiaplane}.}  Since in this limit  the running coupling  $a_s(-s)$ entering (\ref{eq:LaplaceCI}), the fixed scale coupling  $a_s(M_\tau^2)$  
entering (\ref{eq:LaplaceFO}), and the  one-loop coupling $\tilde{a}_s(-s)$ entering (\ref{eq:LaplaceRGS}) are close to each other, it follows that the positions of the 
dominant singularities in the $u$-plane of the corresponding Borel transforms, $B(u)$, $B_{\rm FO}(u,s)$ and $ B_{\rm RGS}(u,y)$,  must be the same.

\section{Analytic continuation in the Borel plane and new perturbative RGS expansions}\label{sec:confmap}
As discussed above, the RGS expansion (\ref{dseriesy}) is divergent, the coefficients $c_{n,1}$  increasing as $n!$ at large $n$. In fact, as we shall show in the next section, the divergence 
is quite bad for expanded functions supposed to resemble the physical Adler function. A procedure to tame this divergent behaviour is therefore mandatory. 
In this section we shall  improve the large-order behaviour of the RGS expansion 
by applying a method based on analytic continuation  in the Borel plane, applied  to the standard CIPT and FOPT in Refs. \cite{CaFi2009,CaFi2011}. 

 The starting remark  is that the Taylor expansion  (\ref{B}) of  $B(u)$  is convergent only inside the disk $|u|<1$, limited by the nearest singularity at $u=-1$.  The region of convergence 
can be enlarged if the series  in powers of  $u$ is replaced by a series in powers of another variable. As shown in \cite{CaFi1998, CaFi2009, CaFi2011},  the ``optimal" variable according to 
the definition proposed in \cite{CiFi} is the function $w\equiv\wt w(u)$ that conformally maps the assumed holomorphy domain of $B(u)$, {\em i.e.} the whole $u$ plane cut along $u\ge 2$ 
and $u\le -1$,   onto the unit disk $|w|<1$ in the  $w$ complex plane. The expansion of $B(u)$ in powers of $w$  is convergent in the whole complex $u$-plane except for the cuts. Moreover, 
this optimal mapping   provides the fastest large-order convergence rate, compared to other variables that conformally map onto the unit disk only parts of the $u$ plane \cite{CiFi}. The 
detailed proof of these statements is given in \cite{CaFi2011}. 

An additional improvement, discussed in detail in \cite{CaFi2011}, is obtained by exploiting the known
 behaviour at the first singularities, presented in  (\ref{eq:gammapowers}) and (\ref{eq:gamma12}). 
The main idea of the procedure, denoted as  singularity softening \cite{SoSu}, is to multiply $B(u)$ with a suitable factor  $S(u)$, such that in the product  $S(u) B(u)$ the 
dominant singularities are compensated  or are replaced by milder singularities.  Moreover,  after compensating the leading singularities, one can  expand the resulting function in powers 
of variables that take  into account only the higher, {\em i.e} more distant, renormalons.

As shown in the previous section, the  Borel transform   $B_{\rm RGS}(u, y)$ of the RGS expansion (\ref{dseriesy}) has the same dominant singularities as $B(u)$.  Therefore we can apply 
the techniques of improving the convergence mentioned above.
 Following \cite{CaFi2011}, we consider the functions 
\beq\label{eq:Wnjk}
\wt w_{lm}(u)=\frac{\sqrt{1+u/l}-\sqrt{1-u/m}}{\sqrt{1+u/l}+\sqrt{1-u/m}},\eeq
 where $l, m$ are positive integers satisfying  $l\ge 1$ and $m \ge 2$.  The function $\wt w_{lm}(u)$ maps the $u$-plane cut along $ u\le -l$ and $u\ge m$ onto the disk $|w_{lm}|<1$ in the plane $w_{lm}\equiv \wt w_{lm}(u)$. The ``optimal" mapping defined above 
is $\wt w(u)\equiv \wt w_{12}(u)$, for which the entire holomorphy domain of the Borel transform is mapped onto the interior of the unit circle in the plane $w_{12}$. 

 We define further the class of compensating factors of the simple  form \cite{CaFi2011}
\beq\label{eq:Sjk}
 S_{lm}(u)=\left(\!1-\frac{\wt w_{lm}(u)}{\wt w_{lm}(-1)}\!\right)^{\!\!\gamma^{(l)}_1} \!\!\left(\!1-\frac{\wt w_{lm}(u)}{\wt w_{lm}(2)}\!\right)^{\!\!\gamma^{(m)}_2}, 
\eeq
 where the exponents, written in terms of the powers $\gamma_l$  defined in (\ref{eq:gammapowers})-(\ref{eq:gamma12}) and the  Kronecker symbol $\delta_{lm}$, as
\beq
\gamma_1^{(l)}=\gamma_1 (1+\delta_{l1}), \quad
 \gamma_2^{(m)}= \gamma_2 (1+\delta_{m2}), 
\eeq
are chosen such that $S_{lm}(u)$ cancel the dominant singularities  defined in (\ref{eq:gammapowers}). Following  \cite{CaFi2011},  we further expand the  product $S_{lm}(u) B_{\rm RGS}(u, y)$  in powers of the variable $\wt w_{lm}(u)$, as 
\beq \label{eq:prod} S_{lm}(u) B_{\rm RGS}(u, y) = 
\sum_{n\ge 0} c_{n, {\rm RGS}}^{(lm)}(y)\, (\wt w_{lm}(u))^n.
\eeq
For the optimal mapping $\wt w_{12}$ this expansion converges in the whole disk $|w_{12}|<1$, {\em i.e.} in the whole $u$-plane cut along $u \ge 2$ and $u\le -1$, and has the best asymptotic convergence rate \cite{CiFi,CaFi2011}. Moreover, since the first singularities of the Borel transform are compensated by the softening factor,  a good convergence is expected  also at finite orders. For other mappings,  with either $l>1$ or $m>2$, the expansions would converge in the unit disks $|w_{lm}|<1$ if the singularities situated between $-l\le u \le -1$ and  $2\le u \le m$ were completely removed by the compensating factors. In practice, however, only the first singularities at $u=-1$ and $u=2$ are compensated by the factors  $S_{lm}(u)$,  and  
after the compensation they may survive  as ``mild" branch points, where the function vanishes 
instead of becoming infinite.  The presence of the residual cuts sets a limit on the convergence domain of the 
expansions (\ref{eq:prod}), but for a mild singularity the effect is expected to become important only at large orders \cite{CaFi2011}.  

While  the optimal mapping is based on a mathematical theorem \cite{CiFi, CaFi2011}, there is no such rigorous result for the form of the softening factors. They are arbitrary to a 
large extent and can be chosen  empirically.  With the choice (\ref{eq:Sjk}), the compensating factors used in different expansions are parametrized in different way. By this we reduce  
the bias related to the choice of these factors.

By combining the expansion (\ref{eq:prod}) with the definition (\ref{eq:LaplaceRGS}), 
we are led to the class of BRGS expansions 
\be\label{eq:DjkRG} \wh D_{\rm BRGS} (s) = \sum\limits_{n\ge 0} c_{n, {\rm RGS}}^{(lm)}(y) \, {\cal W}^{(lm)}_{n,{\rm RGS}} (s),\ee
where
\be\label{eq:WnjkRG} {\cal W}^{(lm)}_{n,{\rm RGS}} (s)=\frac{1}{\beta_0} {\rm PV}
\int\limits_0^\infty\!\exp{\left(\frac{-u}{\beta_0 \tilde a_s(-s)}\right)} \,\frac{(\wt w_{lm}(u))^n}{S_{lm}(u)}\, {\rm d} u,\ee
and the coefficients $c_{n, {\rm RGS}}^{(lm)}(y)$ are defined by the expansion (\ref{eq:prod}).

For completeness we write below the similar ``Borel and contour-improved"  (BCI) expansions \cite{CaFi2009,CaFi2011}
\be\label{eq:DjkCI} \wh D_{\rm BCI} (s) = \sum\limits_{n} c_{n, {\rm CI}}^{(lm)} \, {\cal W}^{(lm)}_{n, {\rm CI}}(s),\ee
where the expansion functions are expressed in terms of the running coupling $a_s(s)$:
\be\label{eq:WnjkCI}   {\cal W}^{(lm)}_{n, {\rm CI}}(s)=\frac{1}{\beta_0} {\rm PV} \int\limits_0^\infty\!{\rm e}^{-u/(\beta_0 a_s(s))} \,  
\frac{(\wt w_{lm}(u))^n}{S_{lm}(u)}\, {\rm d} u,\ee
and the coefficients $c_{n, {\rm CI}}^{(lm)}$ are defined by the expansion
\beq \label{eq:prodCI} S_{lm}(u) B(u) = 
\sum_{n\ge 0} c_{n, {\rm CI}}^{(lm)}\, (\wt w_{lm}(u))^n.
\eeq

Finally, the ``Borel improved fixed-order" (BFO) expansions are written as \cite{CaFi2009,CaFi2011}
 \be\label{eq:DjkFO}  \wh D_{\rm BFO}  (s) = \sum\limits_{n} c_{n, {\rm FO}}^{(lm)}(s) \,{\cal W}^{(lm)}_{n, {\rm FO}},\ee
where the the expansion functions involve the fixed-scale coupling $a_s(M_\tau^2)$:
\be\label{eq:WnjkFO}   {\cal W}^{(lm)}_{n, {\rm FO}}=\frac{1}{\beta_0} {\rm PV} \int\limits_0^\infty\!{\rm e}^{-u/(\beta_0 a_s(M_\tau^2))} \,  
\frac{(\wt w_{lm}(u))^n}{S_{lm}(u)}\, {\rm d} u,\ee
and the coefficients $c_{n, {\rm FO}}^{(lm)}(s)$ are defined by:
\beq \label{eq:prodFO} S_{lm}(u) B_{\rm FO}(u, s) = 
\sum_{n\ge 0} c_{n, {\rm FO}}^{(lm)}(s)\, (\wt w_{lm}(u))^n.
\eeq

The properties of the new expansions  were discussed in detail in \cite{CaFi2000,CaFi2001} in the particular case of the optimal mapping  $ w_{12}$. Their definition is based on a prescription for the  infrared ambiguity of perturbation theory which is consistent with analyticity in the energy plane. 
 When reexpanded in powers of $a_s$,  the expansions reproduce the coefficients $c_{n,1}$   known from Feynman diagrams, up to any order $N$. The remarkable feature is that the expansion  functions ${\cal W}^{(lm)}_n$ resemble the expanded function $\wh D(s)$, being singular at $a_s=0$ and admitting divergent expansions  in powers of the coupling. Therefore, the divergent pattern of the expansion of  $\wh D(s)$ in terms of these new functions is expected to be tamed. The numerical studies reported in the next section confirm this expectation.\footnote{A formal proof  for the optimal mapping  $w_{12}$ was given in \cite{CaFi2000}, where it was shown that under some special conditions the  expansion  (\ref{eq:DjkCI})
is convergent in a domain of the complex $s$-plane.} 
As in \cite{CaFi2011}, we shall use in these analyses the expansions based on the optimal mappings $ w_{12}$ and the alternative mappings $ w_{13}$, $ w_{1\infty}$ and $ w_{23}$.
\section{Models for the Adler functions}\label{sec:tests}
In order to test numerically the convergence properties
of the BRGS expansions defined in the previous section,  we consider a class of physical models  discussed recently in the literature \cite{BeJa, CaFi2009, CaFi2011, DeMa, BeBoJa}.

We first consider the model proposed
in \cite{BeJa}, where the  Adler function $\wh D_{BJ}(s)$ is defined as the {\rm PV}-regulated Laplace-Borel integral
\beq\label{eq:pv}
\wh D_{BJ}(s)=\frac{1}{\beta_0}\,{\rm PV} \,\int\limits_0^\infty  
\exp{\left(\frac{-u}{\beta_0 a_s(-s)}\right)} \, B_{BJ}(u)\, {\rm d} u,
\eeq
in terms of a Borel transform $B_{BJ}(u)$ parametrized in terms of UV and IR renormalons and a regular, polynomial part:
\beq\label{eq:BBJ}
\frac{B_{BJ}(u)}{\pi}=B_1^{\rm UV}(u) +  B_2^{\rm IR}(u) + B_3^{\rm IR}(u) +
d_0^{\rm PO} + d_1^{\rm PO} u.
\eeq
In the expressions of the renormalons 
\bea\label{eq:BIRUV}
B_p^{\rm IR}(u)= 
\frac{d_p^{\rm IR}}{(p-u)^{\gamma_p}}\,
\Big[\, 1 + \tilde b_1 (p-u) + \ldots \,\Big], \nonumber \\
B_p^{\rm UV}(u)=\frac{d_p^{\rm UV}}{(p+u)^{\bar\gamma_p}}\,
\Big[\, 1 + \bar b_1 (p+u)  +\ldots \,\Big],
\eea
 most of the parameters were fixed by imposing renormalization-group 
invariance at four loops. Finally, the free parameters of the model, namely the residues $d_1^{\rm UV}, d_2^{\rm IR}$ and  $d_3^{\rm IR}$ of the first renormalons and the 
coefficients $d_0^{\rm PO}, d_1^{\rm PO}$ of the polynomial in (\ref{eq:BBJ}),  were determined  by the requirement of reproducing 
the perturbative coefficients $c_{n,1}$ for $n\le 4$ from (\ref{cn1}) 
and the estimate $c_{5,1}=283$. Their numerical values are \cite{BeJa}:
\begin{equation}\label{eq:dBJ0}
d_1^{\rm UV}=-1.56\times 10^{-2},~
d_2^{\rm IR}=3.16, ~
d_3^{\rm IR}=-13.5,\\[-3mm]
\end{equation}
\begin{equation}\label{eq:dBJ}
d_0^{\rm PO}=0.781, ~~~
d_1^{\rm PO}=7.66\times 10^{-3}. 
\end{equation}
Then all the higher-order coefficients $c_{n,1}$ are fixed and 
exhibit a factorial increase. The numerical values  up to $n=18$ are listed in \cite{BeJa,CaFi2009}.

The magnitude of the residue $d_2^{\rm IR}$ of the first IR renormalon in the above model was questioned by some authors \cite{Pich_Manchester, DeMa}.  In order to avoid any bias, we have 
also investigated alternative models, in which a smaller residue at $u=2$ was imposed  (an extreme case of this type of alternative models, in which the singularity at $u=2$ is completely 
removed,  was investigated recently in \cite{BeBoJa}). In one example, we have retained the same expressions as in 
\cite{BeJa} for the first three 
singularities and  the same values of the  residues at $u=-1$ and $u=3$, 
while choosing a 
smaller residue at $u=2$, $d_2^{\rm IR}=1$.
 In order to reproduce the first five  coefficients $c_{n,1}$, the model must contain then three additional 
free parameters, which were introduced by a 
quadratic term in the polynomial and two additional IR singularities, at 
$u=4$ and $u=5$. For 
convenience, the nature of these additional singularities, 
which is not known, was taken to 
be the same as that of the $u=3$ singularity. 
Thus, we have considered the alternative model:
\bea\label{eq:altBBJ}
\frac{B_{\rm alt}(u)}{\pi} &=&B_1^{\rm UV}(u) +  B_2^{\rm IR}(u) + B_3^{\rm IR}(u) + 
\frac{d_4^{\rm IR}}{(4-u)^{3.37}}\nn\\
&+&\frac{d_5^{\rm IR}}{(5-u)^{3.37}} + d_0^{\rm PO} + d_1^{\rm PO} u+ d_2^{\rm PO} u^2,
\eea
where, as  discussed above, we have taken as input $d_1^{\rm UV}, 
d_2^{\rm IR}$ and $d_3^{\rm IR}$ from (\ref{eq:dBJ0})
and determined the remaining five parameters by matching the coefficients $c_{n,1}$ for 
$n\le 5$, which gives: 
\be
d_0^{\rm PO}=3.2461, ~~~
d_1^{\rm PO}=1.3680, ~~~ d_2^{\rm PO}=0.2785,\nn\\[-3mm] 
\ee
\be\label{eq:altdBJ1}
d_4^{\rm IR}=560.614, ~~~d_5^{\rm IR}=-1985.73.
\ee

The physical plausibility of this type of models was discussed in several recent works \cite{BeJa,DeMa, Pich_Muenchen, BeBoJa}. In particular, arguments in favor of 
the first model presented above were brought in \cite{BeJa, BeBoJa}.  In the present work  we adopted these models as a mathematical framework 
for testing the  convergence properties of the various expansions.  

\begin{figure}[thb]
 	\begin{center}\vspace{0.5cm}
 	 \includegraphics[width = 6.0cm]{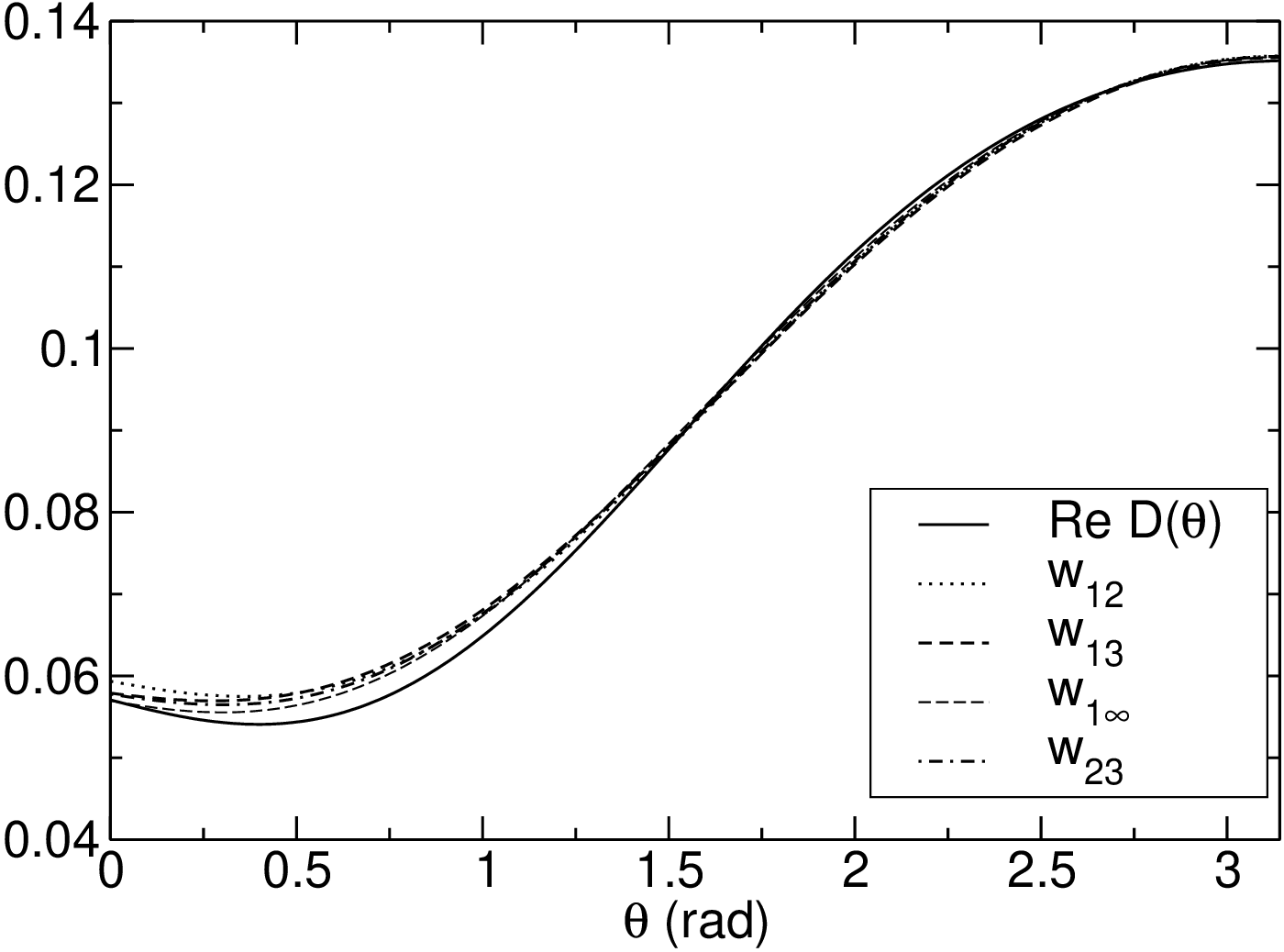}
	\caption{Real part of the Adler function of the model \cite{BeJa} defined 
in (\ref{eq:BBJ})-(\ref{eq:dBJ}), calculated along the circle $s=M_\tau^2 \exp(i\theta)$ 
for $\alpha_s(M_\tau^2)= 0.3156$, using the  BRGS perturbative expansions (\ref{eq:DjkRG})-(\ref{eq:WnjkRG}) with $N=5$ terms. 
The exact function is represented by the solid line.}
	\label{fig:Dre5}
 	\end{center}\vspace{0.0cm}
\end{figure}

\begin{figure}[thb]
 	\begin{center}\vspace{0.5cm}
 	 \includegraphics[width = 6.0cm]{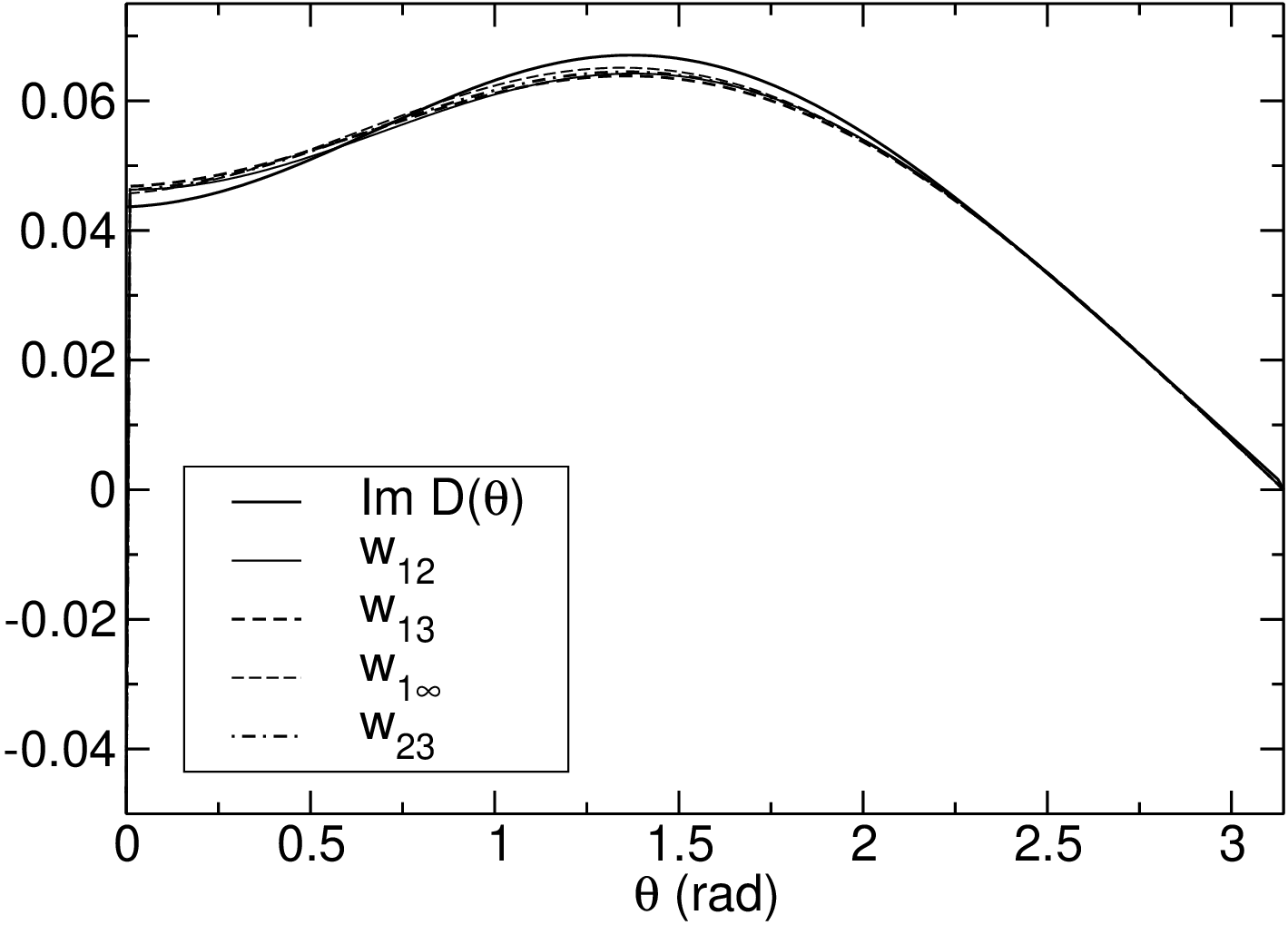}
	\caption{As in Fig. \ref{fig:Dre5} for the imaginary part of the Adler function. }
	\label{fig:Dim5}
 	\end{center}\vspace{0.0cm}
\end{figure}

\begin{figure}[thb]
 	\begin{center}\vspace{0.5cm}
 	 \includegraphics[width = 6.0cm]{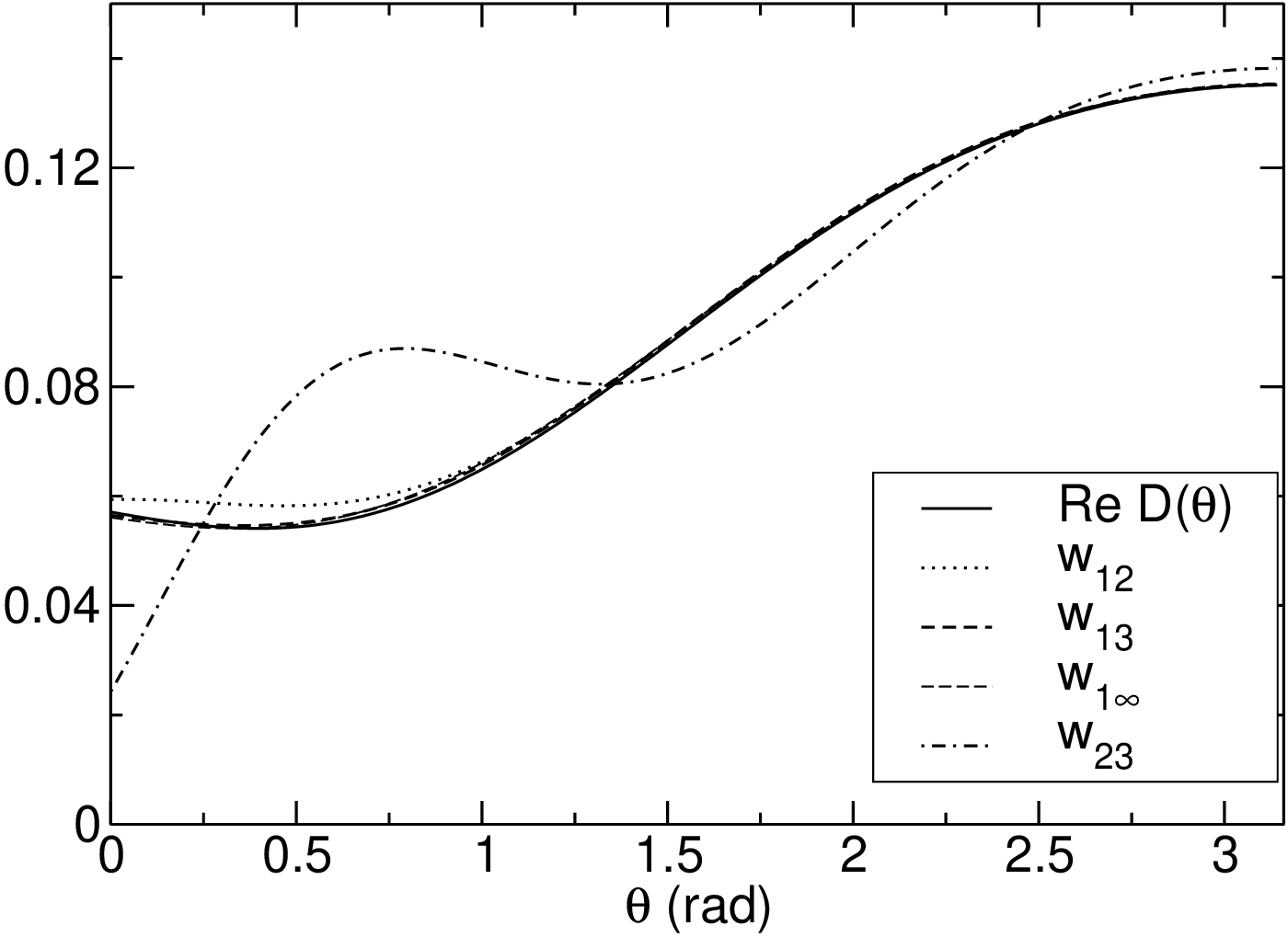}
	\caption{As in Fig. \ref{fig:Dre5} for $N=18$ terms in the expansions.}
	\label{fig:Dre18}
 	\end{center}\vspace{0.0cm}
\end{figure}

\begin{figure}[thb]
 	\begin{center}\vspace{0.5cm}
 	 \includegraphics[width = 6.0cm]{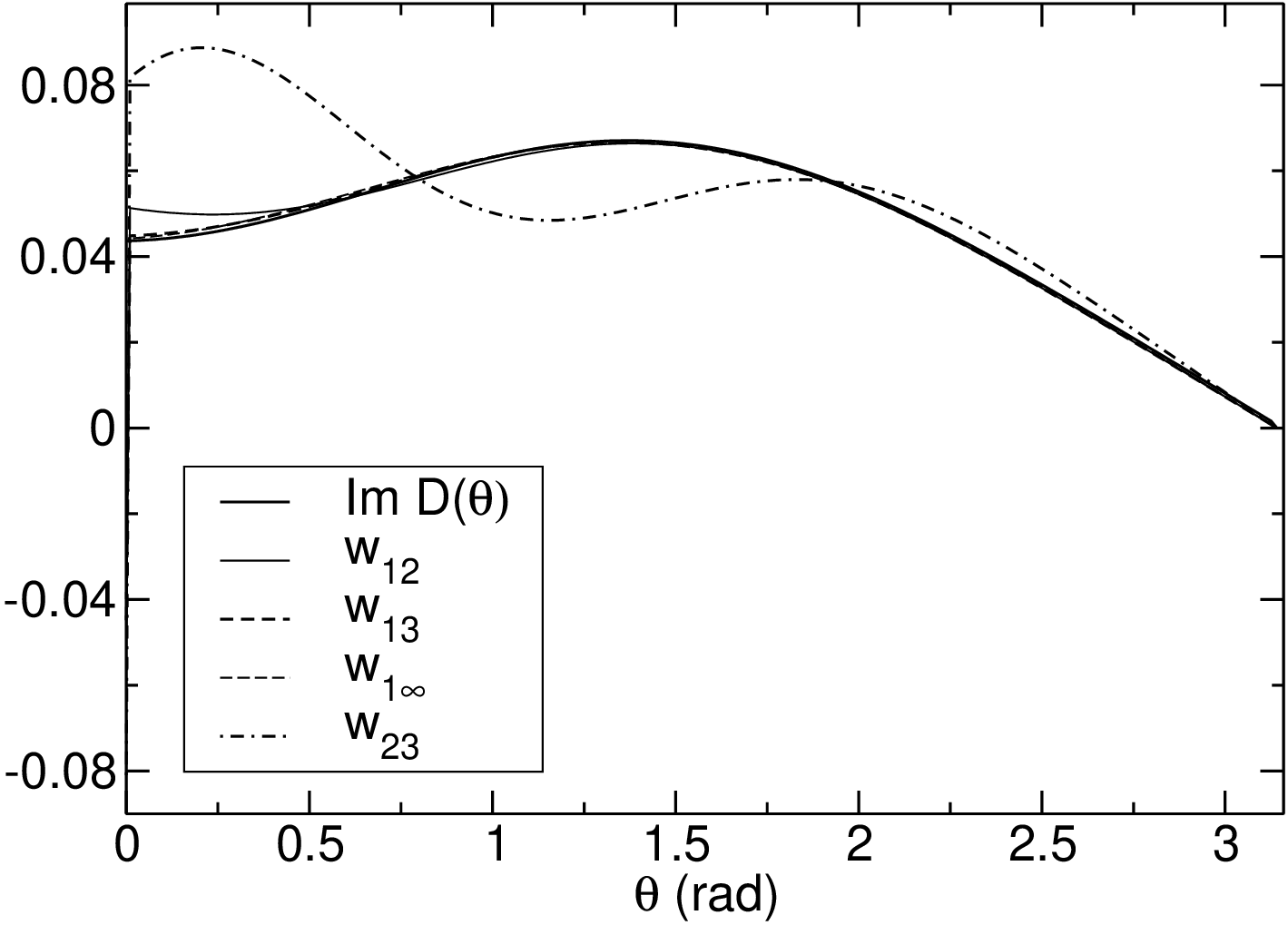}
	\caption{As in Fig. \ref{fig:Dim5} for $N=18$ terms in the expansions. }
	\label{fig:Dim18}
 	\end{center}\vspace{0.0cm}
\end{figure}

We  illustrate first the properties of the new expansions by the approximation they provide to the expanded function  along the circle $s=M_\tau^2 \exp(i\theta)$.  
In Fig. \ref{fig:Dre5}, we show the real part of the Adler function for the model \cite{BeJa} defined 
in (\ref{eq:BBJ})-(\ref{eq:dBJ}), 
calculated  with the new BRGS expansions (\ref{eq:DjkRG})-(\ref{eq:WnjkRG}), with $N=5$ 
terms in the perturbative series. As in \cite{ CaFi2011}, we considered the expansion functions for the optimal mapping
$w_{12}$ and the alternative mappings $w_{13}$, $w_{1\infty}$ and $w_{23}$.
For the
comparison with previous studies
\cite{BeJa, CaFi2009, CaFi2011}, we have used $\alpha_s(M_\tau^2)= 0.3156$ 
in this calculation.
In Fig.\ref{fig:Dim5}, we show the same curves for the imaginary part of the Adler function. 
In Figs. \ref{fig:Dre18} and \ref{fig:Dim18}, 
we repeat our calculations with $N=18$ terms. 

 As shown in Figs. \ref{fig:Dre5}-\ref{fig:Dim18}, the BRGS expansions provide  a good description of the exact function  along the whole circle, including the points near the  
timelike axis, which correspond to
$\theta=0$, and  near the spacelike axis, where $\theta=\pi$.  The worse approximation provided by the mapping $w_{23}$ for  $N=18$ can be explained by the effect of the residual mild 
cut between $u=-1$ and $u=-2$, which limits the convergence radius of the expansion (\ref{eq:prod}) in powers of  $w_{23}$ to $u<1$. For other mappings, the divergence due to the residual 
cuts  is manifest only for $u>2$, and this region is more suppressed by the exponent in the Laplace-Borel integrals (\ref{eq:WnjkRG})  defining the expansion functions   (for more explanations see \cite{CaFi2011}). In all the cases, however, the conformal mappings improve the convergence rate for small values of $u$, which bring the most important  contribution to the Laplace-Borel integral.

Similar representations along the circle $|s|=M_\tau^2$  using the standard CI and FO expansions are given in Refs. \cite{BeJa,CaFi2009}. For the standard RGS expansion the results reported in  \cite{Abbas:2012} show that  its predictions  are quite close to those of the standard CIPT.  The standard expansions exhibit bigger and bigger oscillations with increasing $N$, failing to reproduce with accuracy the exact function along the circle.

The  BRGS expansions should be compared with the  Borel improved CI and FO expansions, defined in (\ref{eq:DjkCI})-(\ref{eq:WnjkCI}) and (\ref{eq:DjkFO})-(\ref{eq:WnjkFO}) 
respectively, which were investigated  in  \cite{CaFi2011} (the particular expansion based on the optimal mapping $w_{12}$ was treated also in \cite{CaFi2009}).  As illustrated in 
the figures given in \cite{CaFi2009, CaFi2011}, the Borel improved CI expansions reproduce very well the exact function, much like the new BRGS expansions, while the Borel improved  FO expansions provide   a very good approximation near the spacelike axis, where the powers of the 
logarithm $\ln(-s/M_\tau^2)$ present in the coefficients are small, but the approximation  becomes worse near the  timelike axis, where  the logarithm acquires a 
large imaginary part.

\begin{table*}[!ht]
\caption{  The difference $\delta^{(0)}-\delta_{\rm exact}^{(0)}$  for the model $B_{\rm BJ}$ proposed 
in \cite{BeJa} 
and specified in (\ref{eq:BBJ})-(\ref{eq:dBJ}), calculated for  
$\alpha_s(M_\tau^2)=0.34$ with the standard CI, FO and RGS expansions, and the new BRGS expansions (\ref{eq:DjkRG})-(\ref{eq:WnjkRG})  for various conformal mappings
 $w_{lm}$, truncated at order $N$. Exact value $\delta_{\rm exact}^{(0)} =0.2371$.} \vspace{0.1cm}
\label{table:1}
\renewcommand{\tabcolsep}{0.9pc} 
\renewcommand{\arraystretch}{1.1} 
\begin{tabular}{lcccccccccc}\hline\hline 
$N$& CI  & FO  & RGS    &BRGS  $w_{12}$ &BRGS $w_{13}$&BRGS $w_{1\infty}$&BRGS $w_{23}$\\\hline 
2&-0.0595&-0.0679&-0.0574&-0.0347& -0.0239& -0.0417& -0.0177\\
3&-0.0473&-0.0345&-0.0440 &-0.0333 & -0.0301& -0.0349& -0.0303\\ 
4&-0.0388&-0.0171&-0.0347&-0.0089& -0.0142& -0.0067& -0.0132 \\ 
5&-0.0349&-0.0083&-0.0315&-0.0070& -0.0086& -0.0058& -0.0070 \\ 
6&-0.0325&-0.0043&-0.0284&-0.0073 & -0.0071& -0.0064& -0.0072\\
7&-0.0325&-0.0029&-0.0298&-0.0059& -0.0057& -0.0056&  -0.0044 \\ 
8&-0.0354&-0.0018&-0.0309&-0.0041& -0.0035& -0.0041&  -0.0011 \\ 
9&-0.0367&-0.0004&-0.0363&-0.0023& -0.0019& -0.0028&  -0.0010\\ 
10&-0.0529&0.0019&-0.0483&0.0014  &  -0.0012& -0.0020&  0.0004\\
11&-0.0409&0.0031&-0.0458&0.0036 & -0.0008& -0.0016&  -0.0009\\
12&-0.1248&0.0065&-0.1335&0.0031 & -0.0006& -0.0015&  0.0005\\ 
13&0.0258&0.0037 &0.0534 &0.0026 &  -0.0004& -0.0015&   -0.0005\\ 
14&-0.5286&0.0204&-0.7850 &0.0018 & -0.0003& -0.0015&   -0.0011\\ 
15&0.8640&-0.0201 &1.7734 &0.0006 &  -0.0002& -0.0015&    0.0044\\ 
16&-3.5991&0.1447&-7.7043& 0.0001& $-7 \cdot 10^{-6}$& -0.0015&    -0.0131\\
17&9.3560&-0.4252 &24.8586& -0.0004& $4 \cdot 10^{-6}$& -0.0014&     0.0238\\ 
18&-31.76&1.907&-94.26&-0.0013& -0.0001& -0.0013&     -0.0310\\ \hline  \hline 
\end{tabular}
\end{table*}

\begin{table*}[!ht]
\caption{ As in Table \ref{table:1}  for the 
modified model $B_{\rm alt}$ specified in (\ref{eq:altBBJ})-(\ref{eq:altdBJ1}). Exact 
value $\delta_{\rm exact}^{(0)}=0.2102$.} \vspace{0.1cm}
\label{table:2}
\renewcommand{\tabcolsep}{0.9pc} 
\renewcommand{\arraystretch}{1.1} 
\begin{tabular}{lcccccccccc}\hline\hline
$N$&  CI  &FO & RGS   & BRGS $w_{12}$ & BRGS $w_{13}$&  BRGS $w_{1\infty}$& BRGS $w_{23}$\\\hline 
2& -0.0326     & -0.0410   & -0.0305  & -0.0078 & 0.0030   & -0.0148  &  0.0092 \\
3&  -0.0204    & -0.0076  & -0.0171  &-0.0064   & -0.0033  & -0.0080  &   -0.0034\\ 
4& -0.0119     & 0.0098   & -0.0078  & 0.0180  & 0.0127   & 0.0202   &   0.0137 \\ 
5& -0.0080      & 0.0186   & -0.0046  & 0.0199  & 0.0183   & 0.0211   &   0.0110  \\ 
6&  -0.0061    & 0.0216   & -0.0026  & 0.0175  & 0.0175   & 0.0182   &   0.0197 \\
7&  -0.0061    & 0.0188   & -0.0047  & 0.0193  & 0.0150   & 0.0153   &   0.0225  \\ 
8&  -0.0079    & 0.0111   & -0.0052  & 0.0201  & 0.0132   & 0.0131   &   0.0258  \\ 
9& -0.0065     & 0.0008   & -0.0078  & 0.0106   & 0.0101   & 0.0109   &   0.0259         \\ 
10& -0.0178    & -0.0070   & -0.0142  & -0.0012 & 0.0047    & 0.0083  &   0.0273 \\
11& 0.0022     & -0.0098  & -0.0020  & -0.0118 & -0.0010   & 0.0058  &   0.0260 \\
12& -0.0690     & -0.0031  & -0.0737  & -0.0231 & -0.0054   & 0.0037  &   0.0274 \\ 
13& 0.1019     & 0.0015   & 0.1397  & -0.0310 & -0.0081   & 0.0023   &   0.0264  \\ 
14& -0.4207    & 0.0242   & -0.6549  & -0.0339 & -0.0093   & 0.0014  &   0.0258  \\ 
15& 1.0234     & -0.0168  & 1.9784  & -0.0347 & -0.0086   & 0.0008   &   0.0313   \\ 
16& -3.3572    & 0.1398   & -7.3731  & -0.0316 & -0.0062   & 0.0004  &   0.0139   \\
17& 9.7378     & -0.4435  & 25.4225  & -0.0239 & -0.0028   & 0.0002  &   0.0507   \\ 
18& -31.15   & 1.874   & -93.316  & -0.0156& 0.0003   & $-2 \cdot 10^{-5}$  &  -0.0041   \\ \hline\hline 
\end{tabular}
\end{table*}

\begin{table*}[!ht]
\caption{  The difference $\delta^{(0)}-\delta_{\rm exact}^{(0)}$  for the model $B_{\rm BJ}$ proposed 
in \cite{BeJa} 
and specified in (\ref{eq:BBJ})-(\ref{eq:dBJ}), calculated for  
$\alpha_s(M_\tau^2)=0.34$ with the  improved BCI expansions  (\ref{eq:DjkCI})-(\ref{eq:WnjkCI}) and the BRGS expansions (\ref{eq:DjkRG})-(\ref{eq:WnjkRG}), for various 
conformal mappings $w_{lm}$, truncated at order $N$. Exact value $\delta_{\rm exact}^{(0)} =0.2371$.} \vspace{0.1cm}
\label{table:3}
\renewcommand{\tabcolsep}{0.9pc} 
\renewcommand{\arraystretch}{1.1} 
\begin{tabular}{lcccccccccc}\hline\hline 
$N$& BCI $w_{12}$ & BRGS  $w_{12}$ &  BCI $w_{13}$& BRGS $w_{13}$ &   BCI $w_{1 \infty}$     &BRGS $w_{1\infty}$ & BCI $w_{23}$&BRGS $w_{23}$\\\hline 
2&   -0.0394    &-0.0347&    -0.0301     &-0.0239&   -0.0488       &-0.0417&      -0.0248             & -0.0177\\
3&  -0.0362     &-0.0333 &    -0.0341    &-0.0301&    -0.0396      &-0.0349&        -0.0343            & -0.0303\\ 
4&   -0.0108    &-0.0089&    -0.0177     & -0.0142&     -0.0083     &-0.0067&      -0.0165            & -0.0132 \\ 
5&    -0.0081   &-0.0070&    -0.0103     &-0.0086&    -0.0061       &-0.0058&       -0.0079             &-0.0070 \\ 
6&   -0.0047     &-0.0073 &  -0.0065      &-0.0071&     -0.005       &-0.0064&        -0.0052             &-0.0072\\
7&  -0.0032     &-0.0059&   -0.004      &-0.0057&      -0.0038        &-0.0056&        -0.0026             &-0.0044 \\ 
8&  -0.0032     &-0.0041&     -0.0028    &-0.0035&     -0.003           &-0.0041&        -0.0024             & -0.0011 \\ 
9&   -0.0030    &-0.0023&   -0.0023      &-0.0019&        -0.0025          &-0.0028&        -0.0024               & -0.0010\\ 
10&   -0.0020  &0.0014  &    -0.0023     &-0.0012&        -0.0023            &-0.0020&       -0.0018                &   0.0004\\
11&  -0.0012   &0.0036 &    -0.0023      &-0.0008&       -0.0022               &-0.0016&      -0.0023                  & -0.0009\\
12&  -0.0009   &0.0031 &   -0.002       &-0.0006&         -0.0022               &-0.0015&         0.0003                &  0.0005\\ 
13&  -0.0009   &0.0026 &   -0.0016       &-0.0004&       -0.0022                    &-0.0015&     -0.0023                    & -0.0005\\ 
14&  -0.0007   &0.0018 &   -0.001       &-0.0003&          -0.0022                    &-0.0015&      0.0024                   &  -0.0011\\ 
15&  -0.0004   &0.0006 &  -0.0005        & -0.0002&         -0.0021                     &-0.0015&     -0.0015                     &  0.0044\\ 
16&  -0.0003   & 0.0001&   -0.0002       &$-7 \cdot 10^{-6}$&        -0.002            & -0.0015&   -0.0028       &-0.0131\\
17&  -0.0003   & -0.0004&  0.0001      & $4 \cdot 10^{-6}$&        -0.0019               &-0.0014&     0.0162        &  0.0238\\ 
18&  -0.0003   &-0.0013&    0.0002     &-0.0001&                    -0.0017                     & -0.0013&      -0.0445                       &  -0.0310\\ \hline  \hline 
\end{tabular}
\end{table*}

\begin{table*}[!ht]
\caption{ As in Table \ref{table:3}  for the 
modified model $B_{\rm alt}$ specified in (\ref{eq:altBBJ})-(\ref{eq:altdBJ1}). Exact  value $\delta_{\rm exact}^{(0)}=0.2102$.} \vspace{0.1cm}
\label{table:4}
\renewcommand{\tabcolsep}{0.9pc} 
\renewcommand{\arraystretch}{1.1} 
\begin{tabular}{lcccccccccc}\hline\hline
$N$&  BCI $w_{12}$  & BRGS $w_{12}$ &    BCI $w_{12}$    & BRGS $w_{13}$&   BCI $w_{1\infty}$  &BRGS $w_{1\infty}$& BCI $w_{23}$   &BRGS $w_{23}$\\\hline 
2&     -0.0125      & -0.0078 &    -0.0032     & 0.0030   &     -0.0219     &-0.0148  &     0.0021      &0.0092 \\
3&   -0.0093        &-0.0064   &    -0.0072     & -0.0033  &    -0.0127     &-0.0080  &    -0.0074       & -0.0034\\ 
4&    0.0161        & 0.0180  &   0.0092       &0.0127   &     0.0186      &0.0202   &     0.0104        &0.0137 \\ 
5&     0.0188      & 0.0199  &     0.0166      &0.0183   &     0.0208       &0.0211   &   0.0190           &0.0110  \\ 
6&     0.0161      & 0.0175  &     0.0169      &0.0175   &    0.0182         &0.0182   &   0.0158           &0.0197 \\
7&    0.0099       & 0.0193  &    0.0118       &0.0150   &    0.0128         &0.0153   &      0.0072         & 0.0225  \\ 
8&   0.0100        & 0.0201  &      0.0062      &0.0132   &     0.0080        &0.0131   &     0.0034        & 0.0258  \\ 
9&   0.0073       & 0.0106   &   0.0041          &0.0101   &   0.0052          &0.0109   &      0.0036         &0.0259    \\ 
10&   -0.0047     & -0.0012 &   0.0042          & 0.0047    &  0.0044          &0.0083  &     0.0013            &0.0273 \\
11&    -0.0120    & -0.0118 &     0.0034        & -0.0010   &    0.0044        &0.0058  &      -0.0034            &0.0260 \\
12&   -0.0095     & -0.0231 &    0.0009         &-0.0054   &   0.0046           &0.0037  &     -0.0021              &0.0274 \\ 
13&   -0.0080     & -0.0310 &     -0.0016        &-0.0081   &    0.0047          &0.0023   &     -0.0042             & 0.0264  \\ 
14&   -0.0101     & -0.0339 &    -0.0028         &-0.0093   &    0.0044           &0.0014  &    0.0022                &0.0258  \\ 
15&   -0.0093     & -0.0347 &     -0.0023        &-0.0086   &    0.0040            &0.0008   &        -0.0015            & 0.0313   \\ 
16&  -0.0058      & -0.0316 &    -0.0011          & -0.0062   &   0.0034            &0.0004  &        -0.0029             &0.0139   \\
17&   -0.0043     & -0.0239 &      0.0       & -0.0028   &    0.0028             &0.0002  &          0.0173            &0.0507   \\ 
18&    -0.0044    & -0.0156&     0.0005         &0.0003   &     0.0022               &$-2 \cdot 10^{-5}$ &     -0.0485             & -0.0041   \\ \hline\hline 
\end{tabular}
\end{table*}

In order to assess the physical relevance of the convergence acceleration of the perturbative expansions, 
we considered the behaviour of the new BRGS schemes 
in the context of $\tau$-lepton hadronic width, which requires the theoretical calculation of the quantity $\delta^{(0)}$ defined in (\ref{del0}). 
In Tables \ref{table:1}-\ref{table:4} we 
give the differences $\delta^{(0)}-\delta^{(0)}_{\rm exact}$
order by order in perturbation theory for the  models discussed above, using various perturbative expansions.  The tendency of this quantity
to flatten out to 0 would indicate that a particular scheme is efficient
and reliable.

In Table \ref{table:1}  we show these differences for the model proposed in \cite{BeJa} and reviewed in Eqs. (\ref{eq:BBJ})-(\ref{eq:dBJ}), and in Table \ref{table:2} we present the results for the alternative model specified in Eqs. (\ref{eq:altBBJ})-(\ref{eq:altdBJ1}). For a consistent comparison with previous results reported in \cite{BeJa, CaFi2009, CaFi2011}, we performed the calculations with
$\alpha_s (M_{\tau}^2) =0.34$.  

 The first three columns of Tables \ref{table:1} and \ref{table:2} show that at low truncation orders the standard FO expansion provides a more precise approximation for the model presented in Table \ref{table:1}, while the standard CI expansion describes better alternative models of the
type shown in Table \ref{table:2}, characterized by a smaller residue of the first IR renormalon. 
These features  were discussed also in \cite{BeJa,BeBoJa}.  At larger orders, the standard FO expansion exhibits in both cases a milder divergence,  explained  \cite{Beneke_Muenchen} by the cancellations between the contributions of the coefficients $c_{n,1}$ and the remaining terms in the series (\ref{DsFO}). 

As concerns  the RGS expansion, it provides at low orders an approximation comparable  to the standard CI expansion for the first model, and slightly better for the second model. However,  the description
 deteriorates beyond $N=10$  where large oscillations in the results appear, the RGS expansion  exhibiting in a more dramatic way than CIPT and FOPT the divergent pattern of the QCD perturbation theory. An improvement of its large-order behaviour by the techniques discussed in this paper is therefore mandatory. 

The last four columns of Table \ref{table:1} show that for the first model the new BRGS expansions provide a very good approximation already at low orders, and the accuracy increases with the truncation order $N$. According to the recent work \cite{BeBoJa}, this model is a solid candidate for the physical Adler function.   For the alternative model, the results shown in Table \ref{table:2} indicate a slightly worse approximation at low orders. However, the good convergence of the new expansions at large orders, in contrast with the big divergencies of the standard expansions, is visible also in this case. 

As we mentioned, the standard RGS expansion is rather similar to the standard CI expansion up to relatively large orders, beyond which  the RGS expansion starts to exhibit much wilder oscillations. It is of interest to compare these schemes also in the Borel improved versions given in Eqs.  (\ref{eq:DjkRG})-(\ref{eq:WnjkRG}) and  (\ref{eq:DjkCI})-(\ref{eq:WnjkCI}), respectively. This comparison is presented in Tables   \ref{table:3} and  \ref{table:4}. The results show that, with small variations, the approximation provided by the BCI and BRGS expansions is similar up to large truncation orders $N$, for both models considered. Thus, using the  technique of series acceleration by conformal mappings of the Borel plane,  the strong divergence of the standard RGS was considerably tamed.  We recall that an advantage of the BRGS expansion is that it does not require the numerical determination of the running coupling along the integration circle $|s|=M_\tau^2$,  involving only analytical expressions.

\section{Determination of $\alpha_s(M_\tau^2)$}\label{sec:alphas}
In this section we shall use the new BRGS expansions defined in the present paper for a new 
determination of $\alpha_s(M_\tau^2)$ in the $\overline{\rm MS}$ scheme.
The determination of this fundamental parameter is one of
the important goals of this work and significant care has to
be exercised in adopting proper values of input along with the experimental
and theoretical uncertainties.  

 We use as input the recent phenomenological value of the pure perturbative correction 
to the hadronic $\tau$ width  \cite{Beneke_Muenchen}
\begin{equation}\label{delph}
 \delta^{(0)}_{\rm phen}=0.2037\pm 0.0040_{\rm exp}\pm 0.0037_{\rm PC},
\end{equation}
where the first error is experimental and the second reflects the uncertainty of the higher-order terms (``power corrections") in the OPE estimated by reasonable theoretical assumptions.  
We emphasize that our calculation 
is not based on the models discussed in the previous section, but relies only on the 
known coefficients $c_{n,1}$ given in (\ref{cn1}),  and the conservative choice  $c_{5,1}=283\pm 283$ 
 for the next coefficient \cite{BeJa, Beneke_Muenchen}.

Using this input, the values of  $\alpha_s(M_\tau^2)$ obtained with the  new BRGS expansions 
defined in (\ref{eq:DjkRG})-(\ref{eq:WnjkRG}), with the expansion functions ${\cal W}_{n, {\rm RGS}}^{(12)}$,   ${\cal W}_{n, {\rm RGS}}^{(13)}$,  ${\cal W}_{n, {\rm RGS}}^{(1\infty)}$ and  ${\cal W}_{n, {\rm RGS}}^{(23)}$, respectively, are:
\begin{eqnarray}\label{eq:alphas}
&&\hspace{-0.5cm} 0.3189 \pm 0.0034_{\rm exp}\pm 0.0031_{\rm PC} 
~ ^{+0.0162}_{-0.0121}(\rm c_{51})~^{+0.0014}_{-0.0013}{(\rm \beta_4)}  ,\nonumber\\
&&\hspace{-0.5cm} 0.3198\pm 0.0034_{\rm exp} \pm 0.0031_{\rm PC} 
~ ^{+0.0112}_{-0.0088}(\rm c_{51})~^{+0.0007}_{-0.0007}{(\rm \beta_4)}   ,\nonumber\\
&&\hspace{-0.5cm} 0.3180\pm 0.0034_{\rm exp} \pm 0.0031_{\rm PC} 
~ ^{+0.0134}_{-0.0103}(\rm c_{51})~^{+0.0010}_{-0.0009}{(\rm \beta_4)},    \nonumber\\
&&\hspace{-0.5cm} 0.3188 \pm 0.0034_{\rm exp} \pm 0.0031_{\rm PC}
~ ^{+0.0143}_{-0.0107}(\rm c_{51})~^{+0.0010}_{-0.0009}{(\rm \beta_4)} . \nonumber\\
\end{eqnarray}

The first two errors are due to the uncertainties of the phenomenological value of $\delta^{(0)}$ given in (\ref{delph}).  The third one,  produced by
the range adopted for the coefficient $c_{5,1}$,  brings the most important contribution to the total error.  The last uncertainty accounts for the higher terms in the 
expansion of the $\beta$ function, simulated as in \cite{Pich_Manchester} by an additional coefficient $\beta_4 = \pm \beta_{3}^2/\beta_2$ in this expansion. We explored also the influence of the scale variation,  choosing it as $\mu^2 = \xi M_{\tau}^2$ with $\xi= 1\pm 0.5$, but the effects are very small, so we show only the result corresponding to the choice $\mu^2 =  M_{\tau}^2$ in (\ref{atilde}).

A very small sensitivity of $\alpha_s(M_\tau^2)$ to the variation of the scale is specific also to the standard CIPT analyses \cite{Pich_Muenchen, Pich_Manchester}, the 
RGS expansion \cite{Abbas:2012}, and the Borel improved CI expansions  \cite{CaFi2009,CaFi2011}. The uncertainty related to the  coefficient  $c_{5,1}$ is bigger in the case of 
the Borel improved expansions than in the standard CI and RGS.  However, as discussed in   \cite{CaFi2009,CaFi2011}, having in view the divergent character of the series, the truncation 
error in the latter versions is certainly underestimated. The calculation of the five-loop coefficient  $c_{5,1}$ is therefore of great interest, as it  would reduce  considerably the 
total error of $\alpha_s(M_\tau^2)$ determined from the Borel improved perturbation schemes. 

By taking the average of the central values and of the errors given in Eq. (\ref{eq:alphas}) we obtain the prediction
\bea\label{eq:alpha1}
 \alpha_s(M_\tau^2)&=& 0.3189 \pm 0.0034_{\rm exp} \pm 0.0031_{\rm PC}\\
&& ^{+0.0138}_{-0.0105}(\rm c_{5,1})~\pm 0.0010_{\rm \beta_4},\nonumber
\eea
which becomes, after adding  the errors in quadrature,
\begin{equation}\label{eq:alpha2}
 \alpha_s(M_\tau^2)= 0.3189~^{+ 0.0145}_{-0.0115 }\,.
\end{equation}

We emphasize that the error quoted above was obtained as the average of the errors of the individual determinations (\ref{eq:alphas}).  Much lower uncertainties would have been obtained if standard statistical  procedures for combining independent determinations  were applied. In practice, although the values given in (\ref{eq:alphas}) may be considered independent theoretical determinations, we prefer the conservative errors given in (\ref{eq:alpha1}), which avoid any bias. We note however the remarkable consistency of the  theoretical determinations given in (\ref{eq:alphas}), which is  a strong  argument in favor of our predictions. It is remarkable also that the central value of our prediction (\ref{eq:alpha2})  practically coincides with the world average  $\alpha_s(M_\tau^2)= 0.3186 \pm 0.0056$ \cite{RPP}.

By evolving  (\ref{eq:alpha2}) to the scale of $M_Z$,  using the CRunDec package \cite{Schmidt:2012az},  our prediction reads
\begin{equation}\label{eq:alpha2Z}
 \alpha_s(M_Z^2)= 0.1184~^{+0.0018}_{-0.0015}\,,
\end{equation}
where the central value coincides with the 2012 world average,  $\alpha_s(M_Z^2)=0.1184 \pm 0.0007 $ \cite{RPP}.

It is of interest to compare the result (\ref{eq:alpha2}) obtained with the BRGS expansions defined in the present work with  other recent determinations of  $\alpha_s(M_\tau^2)$.  The values reported in the recent works \cite{Davier2008}-\cite{Abbas:2012}  are not all based on the same  input. Therefore, for a consistent comparison of different  perturbative schemes, we use in what follows the same input for the phenomenological value of $\delta_{\rm phen}^{(0)}$, given in (\ref{delph}), and the five-loop coefficient, $c_{5,1}=283\pm 283$.
Then the standard FO, CI and RGS expansions lead to the predictions 
\bea\label{FOCIRG}
\alpha_s(M_\tau^2)&=& 0.3199~^{+0.0118}_{-0.0074} \quad\quad\quad {\rm FO}\,, ~~\nonumber \\ 
\alpha_s(M_\tau^2)&=& 0.3419~^{+0.0084}_{-0.0085} \quad\quad \quad {\rm CI}\,, \nonumber \\ 
\alpha_s(M_\tau^2)&= & 0.3378~^{+0.0088}_{-0.0095} \quad\quad\quad {\rm RGS}\,, ~~
\eea
where the FO result is quoted in \cite{Beneke_Muenchen} and the RGS one in \cite{Abbas:2012}. The FO and CI expansions improved by conformal mappings of the Borel plane (denoted as BFO and BCI, respectively) give \cite{CaFi2009,CaFi2011}
\bea\label{FOCI1}
\alpha_{s}(M_{\tau}^2)&=& 0.3109~^{+0.0114}_{-0.0049} \quad\quad\quad{\rm BFO}\,, \nonumber \\
\alpha_{s}(M_{\tau}^2)&=&  0.3195~^{+0.0189}_{-0.0138}\quad\quad\quad{\rm BCI}\,.
\eea 

One can see that the BRGS and BCI expansions, which implement both Borel and RG-summation, lead to similar values for $\alpha_{s}(M_{\tau}^2)$, which actually are also close to the 
prediction of standard FOPT, where neither of the two summations is performed. On the other hand, the standard CI and RGS expansions, where only the summation related to RG is implemented, 
lead to values larger by about 0.02, 
while the  BFO expansions, 
which improve only the large-order behaviour without summing the terms known from  the RG, lead to a value  smaller by about 0.01 than the predictions of  BRGS, BCI and FOPT.

In the recent work, Ref. \cite{Boito_update}, it has been pointed out,
on the basis of a detailed analysis of moments of the spectral functions measured by OPAL experiment, that
the uncertainty of the nonperturbative contributions to the hadronic $\tau$-width may be larger than usually assumed.
To account for this possibility, we adopt a more conservative value for the uncertainty of $\delta^{(0)}_{\rm phen}$ due to power corrections, {\it viz.}  
replacing  $\pm 0.0037$ in (\ref{delph}) by the estimate $\pm 0.012$ of this uncertainty quoted in \cite{Boito_update}.

Using this input, the value $\pm 0.0031$ of the error on $\alpha_s(M_\tau^2)$ produced by the power corrections, quoted in (\ref{eq:alpha1}), increases to  $^{+0.0099}_{-0.0103}$\,, leading to 
\begin{equation}\label{eq:alpha2_update}
 \alpha_s(M_\tau^2)= 0.3189~^{+ 0.0173}_{-0.0151}\,.
\end{equation}
We shall adopt this result, having the same central value and a more conservative error  than in (\ref{eq:alpha2}), as our final prediction.

By evolving  (\ref{eq:alpha2_update}) to the scale of $M_Z$ our prediction reads
\begin{equation}\label{eq:alpha2Z_update}
 \alpha_s(M_Z^2)= 0.1184~^{+0.0021}_{-0.0018}\,.
\end{equation}
\section{Discussion and conclusions}\label{sec:conc}
 The non-strange hadronic decays of the $\tau$ lepton provide one of the most important ways of extracting the strong coupling $\alpha_s$. The
perturbative schemes of the Adler function  that are used in this extraction continue, however,
to be a significant source of uncertainty.  There are two major ambiguities that affect the QCD perturbation expansions: one is related to the implementation of the renormalization-group 
invariance, the other regards the large-order behaviour of the series.

The first ambiguity is usually illustrated by the significant difference between  the predictions of the FO and  CI expansions  of the Adler function. 
In a recent work, Ref. \cite{Abbas:2012}, we  used for the analysis of the $\tau$ hadronic width  another renormalization-group-improved expansion, proposed  in \cite{Ahmady1,Ahmady2}, 
which  
sums in an analytically closed form the logarithms accessible by RG invariance.

The second ambiguity  is associated with the fact that the perturbative series are divergent, due to a factorial growth of the coefficients calculated from Feynman 
diagrams \cite{Muell,Muell1,Broa,Bene}.  In QCD this ambiguity  is even more dramatic than in QED  due to the  IR renormalons, singularities in the Borel 
plane situated on the positive axis, which prevent the unique reconstruction of the function by means of a Laplace-Borel integral. However, if 
one adopts a suitable prescription (principal value), it is possible to  exploit the available knowledge on the large-order behaviour of the coefficients for defining a new expansion, in 
which the 
divergent pattern is tamed to a great extent.
Such a method was proposed some time ago  in \cite{CaFi1998,CaFi2000,CaFi2001} and was applied recently
in \cite{CaFi2009,CaFi_Manchester,CaFi2011} to both the FO and CI expansions of the Adler function. It is based on series acceleration by analytic continuation in the Borel plane, achieved by conformal mappings and the ``softening" of the dominant singularities of the Borel transform.  

In the present work, we used this method in order to improve the large-order behaviour of the RGS expansion discussed in \cite{Abbas:2012}. That it should be possible to have a 
straightforward application of the 
technique is not obvious, as the RGS expansion 
(\ref{dseries}) involves a set of complicated functions $D_n(y)$ determined iteratively by the differential equations (\ref{Dk_de}). However, after expressing the series in the 
alternative form (\ref{dseriesy}),{\it i.e.,} as an expansion in powers of the one-loop coupling (\ref{atilde}), it was possible to show in Sec. \ref{sec:Borel} that the dominant 
singularities of this expansion in the Borel plane  coincide with those of the standard Borel transform $B(u)$ of the CI expansion.
As a result, we are able to apply the techniques discussed in   \cite{CaFi2009,CaFi_Manchester,CaFi2011}, defining  the class of improved 
expansions (\ref{eq:DjkRG})-(\ref{eq:WnjkRG}), denoted as BRGS expansions.  

As discussed in earlier works \cite{CaFi2001,CaFi2011}, the  expansions based on conformal mappings of the Borel plane have a number of remarkable properties. In particular, 
the divergent pattern of the expansion of  $\wh D(s)$ in terms of these new functions is expected to be tamed.
The detailed numerical studies of two representative  models for the Adler function, presented in Sec. \ref{sec:tests},   show that indeed the  expansions  improved by both 
RG summation and analytic continuation in the Borel plane, {\em i.e.} BCI and BRGS, provide a very good  approximation of the true functions in the complex energy plane up to 
high orders. This is seen from Figs. \ref{fig:Dre5}-\ref{fig:Dim18} of Sec. \ref{sec:tests} and the similar figures presented in Refs.  \cite{CaFi2009, CaFi2011}. As a consequence, 
as shown in Tables \ref{table:1} - \ref{table:4}, the approximation of the integral  $\delta^{(0)}$ provided by these expansions is very good and improves with increasing $N$. In 
contrast, the standard expansions   shown in the first three columns of  Tables \ref{table:1} and \ref{table:2} exhibit a divergent pattern.

 As argued in \cite{BeJa,Beneke_Muenchen} and is seen also from our results,  the  FO expansion of the observable  $\delta^{(0)}$  exhibits   a  better behaviour at high orders,  which results from suitable cancellations between the increasing coefficients  $c_{n,1}$ and the remaining terms in (\ref{DsFO}). This cancellation is spoiled in the standard CI expansion (\ref{DsCI}), which sums only the RG part, leaving the increase of  $c_{n,1}$ uncompensated. The better large-order behaviour of FOPT is one of the arguments used in favor of this expansion compared  to CIPT \cite{Beneke_Muenchen}.   As concerns the standard RGS expansion, some cancellations are taking place at low orders, but above $N=10$ this expansion exhibits a divergence pattern even more dramatic than CIPT.  

It should be noted however that the better behaviour of FOPT is restricted only to integrals like $\delta^{(0)}$  or some special moments of the spectral function: as shown in previous studies of specific models \cite{BeJa,CaFi2009, CaFi2011}, the pointwise
description of the true Adler function along the circle $|s|=M_\tau^2$ provided by the FO expansion is quite poor.  The good convergence obtained for some integrals along the circle  is due to fortuitous cancellations of large contributions from different integration regions. On the other hand,  as we mentioned, the BCI and BRGS expansions yield a good pointwise approximation of the true function along the whole circle.

For the determination  of $\alpha_s(M_\tau^2)$  presented in Sec. \ref{sec:alphas} we used as input the phenomenological value of the QCD correction $\delta^{(0)}$ to the $\tau$ hadronic width  evaluated  in  \cite{Beneke_Muenchen}. Our analysis
shows that the   BRGS  and BCI expansions, improved by both RG and high-order Borel summation, lead to similar  results for  $\alpha_s(M_\tau^2)$, which are actually also very close to the standard FOPT prediction obtained with the same input.
It is remarkable that the central value of the prediction (\ref{eq:alpha2})  obtained with the new BRGS expansions  coincides with the world average quoted in \cite{RPP}. We also considered the possibility that the uncertainty related to
nonperturbative contributions  might be larger, as follows from the recent analysis \cite{Boito_update}.  The  error of our final prediction (\ref{eq:alpha2_update}) is a conservative estimate that takes into account this possibility. 

In conclusion, we advocate the use of the BCI and BRGS expansions of the QCD Adler functions, which implement simultaneously the RG invariance and the available knowledge about the large-order 
behaviour of perturbation theory. In particular,  the BRGS expansions proposed in the present work have the advantage  that RG summation is implemented through analytically closed 
expressions. Therefore, these expansions are suitable for more detailed analyses of $\tau$ hadronic decays, based on the  moments of the hadronic spectral function.

\vspace{-0.cm}

\subsection*{Acknowledgements} IC acknowledges support from ANCS, under Contract PN 09370102/2009, and from CNCS in the Program Idei-PCE, Contract No 121/2011. The work was also supported by the Projects No. LA08015 of the Ministry of Education and AV0-Z10100502 of the Academy of Sciences of the Czech Republic.

\end{document}